\documentclass[aps,prd,nofootinbib,preprintnumbers,nofootinbibn,superscriptaddress]{revtex4}
\usepackage{graphicx}  
\usepackage{dcolumn}   
\usepackage{bm}        
\usepackage{amssymb}   
\usepackage[utf8]{inputenc}
\usepackage[english]{babel}
\usepackage{shapepar}

\usepackage{amsmath}
\usepackage{amsthm}
\usepackage{array}
\usepackage{fancybox}
\usepackage{multirow}
\usepackage{appendix}
\usepackage{epsfig}
\usepackage{cases}
\usepackage{forest}
\usepackage{stackrel}

\usepackage{graphicx}
\usepackage{dcolumn}
\usepackage{bm}
\usepackage{color}
\RequirePackage[colorlinks=true
,urlcolor=blue
,anchorcolor=blue
,citecolor=blue
,filecolor=blue
,linkcolor=blue
,menucolor=blue
,linktocpage=true
,pdfproducer=medialab
,pdfa=true
]{hyperref}

\theoremstyle{definition}

\begin{document}


\title{Ghost-free higher-order theories of gravity with torsion}

\author{\'Alvaro de la Cruz-Dombriz}
\affiliation{Cosmology and Gravity Group, Department of Mathematics and Applied Mathematics, University of Cape Town, Rondebosch 7701, Cape Town, South Africa.}
\author{Francisco Jos\'e Maldonado Torralba}  \email{f.j.maldonado.torralba@rug.nl}
\affiliation{Cosmology and Gravity Group, Department of Mathematics and Applied Mathematics, University of Cape Town, Rondebosch 7701, Cape Town, South Africa.}
\affiliation{Van Swinderen Institute, University of Groningen, 9747 AG Groningen, The Netherlands.}
\author{Anupam Mazumdar}
\affiliation{Van Swinderen Institute, University of Groningen, 9747 AG Groningen, The Netherlands.}

\begin{abstract}
In this manuscript we will present the theoretical framework of the recently proposed infinite derivative theory of gravity with a non-symmetric connection. We will explicitly derive the field equations at the linear level and obtain new solutions with a non-trivial form of the torsion tensor in the presence of a fermionic source, and show that these solutions are both ghost and singularity-free.
\end{abstract}

\maketitle

\section{Introduction}

The theory of General Relativity (GR), one of the most celebrated theories in physics, has passed all the experimental tests so far \cite{Will:2014kxa}, including large-scale structure formation~\cite{Dayal:2018hft}, and recent detection of gravitational waves from binary compact objects~\cite{Abbott:2016blz}. Nevertheless, at a classical level, the introduction of fermionic matter in the energy-momentum tensor requires a new formalism to be developed in order to take into account of how internal spin degrees of freedom affect the geometry~\cite{Hehl:1994ue}. This can be addressed by incorporating the gauge structure of the Poincar\'e group, provided a torsion field is added, for a review see~\cite{Gauge,Shapiro:2001rz}, as was first shown by Sciama and Kibble in~\cite{sciama1962analogy} and~\cite{Kibble:1961ba}, respectively. Following this approach one arrives to the conclusion that the space-time connection must be metric compatible, albeit not necessarily symmetric. Hence, a non vanishing torsion field $T_{\,\,\nu\rho}^{\mu}$ (where $\mu,\nu, \rho=0,1,2,3$) ought to be present as a consequence of non-symmetric character of the connection. An interesting fact about these theories is that they arise naturally as {\it gauge theories of Poincar\'e group}, rendering their mathematical formalism analogous to the one used in the Standard Model of particles, and therefore making them good candidates to be explored for the quantization of gravity, and making gravity and gauge theories at par. However, both GR and Poincar\'e gauge gravity suffer from the short distance behavior at a classical level, which results in black-hole-like and cosmological Big Bang singularities \cite{Hawking:1973uf,Cembranos:2016xqx}, known as the ultraviolet (UV) problem. 

The aim of this article will be to provide the foundations of a theory proposed by the authors in \cite{delaCruz-Dombriz:2018aal}. Such a theory is able to recover GR and Poincar\'e gauge theories of gravity in the infrared (IR), while ameliorating the UV behavior, i.e., being singularity-free, in both metric and torsion fields, and also free from ghosts at the linear level. This has been made possible thanks to the introduction of infinite derivatives in the action, that can indeed potentially ameliorate the classical UV behavior, weakening the $1/r$ potential of the weak field limit. In the torsion-free limit such theories have been explored widely, and are known as infinite derivative theories of gravity (IDG), which can be made devoid of ghosts and singularities. The most general action has been constructed around Minkowski spacetime~\cite{Biswas:2011ar}, and in de Sitter and anti-de Sitter~\cite{Biswas:2016etb}. It was shown that the graviton propagator of such theories can be modified to avoid any ghosts, expressing the non-local functions as exponentials of an {\it entire function}, which does not introduce any new complex poles, nor any new dynamical degrees of freedom. Therefore, such theories retain the original 2 dynamical degrees of freedom of GR, i.e., a transverse traceless graviton. However, being infinite derivative theories, such an action introduces non-local gravitational interaction and has been argued to improve UV aspects of quantum nature of gravity~\cite{Tomboulis:1997gg,Modesto:2011kw}, due to the fact that infinite derivatives in general do not have a point support.  IDG theories are also motivated from particle limit of strings~\cite{Tseytlin:1995uq,Siegel:2003vt,Abel:2019ufz}, and have been argued to have no singular static solution. At a quantum level, infinite derivative field theories, with similar propagator, bring softening of UV divergences~\cite{Efimov:1967pjn,Efimov:1968flw,Biswas:2014yia,Tomboulis:2015gfa,Ghoshal:2017egr,Buoninfante:2018mre,Talaganis:2014ida,Talaganis:2016ovm} such as shifting the scale of non-locality with large number of external legs interacting non-locally~\cite{Buoninfante:2018gce}.

At a classical level, it has been shown that such IDG theories can yield a non-singular, static solution at the full non-linear level~\cite{Buoninfante:2018rlq}, can avoid ring singularities in a rotating metric at the linear level~\cite{Buoninfante:2018xif}, and also resolve charged source singularity at the linear level~\cite{Buoninfante:2018stt}. At a dynamical level such theories do not give rise to formation of a trapped surface~\cite{Frolov:2015bta,Frolov:2015bia,Frolov:2015usa}, and possibly even at the level of astrophysical masses there may not posses an event horizon~\cite{Koshelev:2017bxd,Buoninfante:2019swn}. Exact solutions for IDG have been found in~\cite{Biswas:2005qr,Buoninfante:2018rlq,Kilicarslan:2018yxd}, including static and time dependent solutions.


Motivated by these studies of IDG, various extensions of IDG have been made in the context of {\it teleparallel} gravity \cite{Koivisto:2018loq} and symmetric {\it teleparallel} gravity \cite{Conroy:2017yln}, as well as in what regards the extension of Poincar\'e gauge gravity that is well-behaved at the UV at a classical level  \cite{delaCruz-Dombriz:2018aal}. In this article we will give more insights into the theory with explicit calculations, obtaining the classical limit and providing new solutions.  In a previous communication \cite{delaCruz-Dombriz:2018aal}, the authors provided a solution of the IDG with a non-symmetric connection in the presence of an uncharged fermionic source. In this particular case, only the totally-antisymmetric part of the torsion was considered to be non-null, therefore Einstein and Cartan equations became decoupled and the solutions of the metric were the same as those of in metric IDG case~\cite{Biswas:2011ar}. In this work, we will show that having a charged fermion as a source, and assuming also the trace of the torsion to be different from zero, we can find new solutions with a non-trivial torsion tensor. The rest of the components of the torsion tensor have been considered to be negligible motivated by the fact that in the standard cosmological scenario, i.e. having a Friedmann-Lemaitre-Robertson-Walker (FLRW) metric, this component is identically zero \cite{Sur:2013aia}.\\

We have divided the manuscript in the following sections. In Section \ref{sec:II} we will provide the action of the theory and its linearised form. In Section \ref{sec:III} we will derive the field equations of the linearized theory, performing variations with respect to both metric and contorsion. In Section \ref{sec:IV}, we will decompose the contorsion tensor into its three irreducible Lorentz invariants, and rewrite the field equations in terms of them. In Section \ref{sec:V}, we will provide new solutions of the linearized field equations in presence of a fermionic source. Finally, in Section \ref{sec:VI}, we will conclude our analysis with future outlook. 
For the sake of simplicity, we have written the calculations to obtain the linearized Lagrangian of the theory in the Appendices \ref{ap1} and \ref{ap2}. Moreover, in Appendix \ref{ap3} we calculate the local limit of the theory and show that we can recover a local Poincar\'e gauge gravity.

\section{The linearised action}
\label{sec:II}

In the standard IDG theories the connection is metric and symmetric, i.e., the Levi-Civita one. Therefore, the linear action of IDG is built with the gravitational invariants and derivatives, considering only up to order $\mathcal{O}(h^{2})$, where $h$ is the linear pertubation around the Minkowski metric~\footnote{We can go beyond quadratic order in curvature, but of the metric potentials are less than unity then higher curvature effects such as cubic, quadratic, and quintic order curvature terms would be even smaller. If however, there are solutions at the quadratic curvature which become non-linear then we would have to explore higher curvature effects even at the classical level.}
\begin{equation}
g_{\mu\nu}=\eta_{\mu\nu}+h_{\mu\nu}.
\end{equation}
After substituting the linear expressions of the curvature tensors (Riemann, Ricci and curvature scalar) we can obtain the linearised action as first was shown in Ref. \cite{Biswas:2011ar}. With this in mind, we wish to generalize the expressions of the curvature tensor when we consider non-symmetric connection. First, we must take into account that the torsion tensor is not geometrically related to the metric, therefore the conditions that are imposed in $h_{\mu\nu}$ are not enough to construct the linear action in connection. In order to so lve this problem, we will have to impose that the total connection must be of order $\mathcal{O}(h)$, i.e., the same as the Levi-Civita one~\footnote{If this were not the case, we would have two options: either the contribution of the metric is of higher-order than the torsion, hence recovering the usual IDG theory ~\cite{Biswas:2011ar}, or the torsion is of higher order than the metric. In the latter case we would have a somewhat similar action of the UV extension of telleparalel gravity~\cite{Koivisto:2018loq}.}. Then, by using the relation between the Levi-Civita $\Gamma$, and the total connection $\widetilde{\Gamma}$, we can write
\begin{equation}
\widetilde{\Gamma}^{\rho}_{\,\,\mu\nu}=\Gamma^{\rho}_{\,\,\mu\nu}+K^{\rho}_{\,\,\mu\nu},
\end{equation}
where the so-called contortion tensor $K$ must be of the same order as the metric perturbation. This may seem as a strong assumption, nevertheless, as it has been known in the literature, the current constraints on torsion suggest that its influence is very small compared to the purely metric gravitational effects \cite{Lammerzahl:1997wk,Kostelecky:2007kx}. Therefore, considering a higher order perturbation in the torsion sector would make no sense physically.

The way to generalize  the  IDG action will be to consider all the quadratic Lorentz invariant terms that can be constructed with the curvature tensors, the contorsion, and infinite derivatives operators, namely~\cite{delaCruz-Dombriz:2018aal}
\begin{equation}\label{GEQ}
S=\int {\rm d}^{4}x\sqrt{-g}\left[\frac{\widetilde{R}}{2}+\widetilde{R}_{\mu_{1}\nu_{1}\rho_{1}\sigma_{1}}\mathcal{O}_{\mu_{2}\nu_{2}\rho_{2}\sigma_{2}}^{\mu_{1}\nu_{1}\rho_{1}\sigma_{1}}\widetilde{R}^{\mu_{2}\nu_{2}\rho_{2}\sigma_{2}}+\widetilde{R}_{\mu_{1}\nu_{1}\rho_{1}\sigma_{1}}\mathcal{O}_{\mu_{2}\nu_{2}\rho_{2}}^{\mu_{1}\nu_{1}\rho_{1}\sigma_{1}}K^{\mu_{2}\nu_{2}\rho_{2}}+K_{\mu_{1}\nu_{1}\rho_{1}}\mathcal{O}_{\mu_{2}\nu_{2}\rho_{2}}^{\mu_{1}\nu_{1}\rho_{1}}K^{\mu_{2}\nu_{2}\rho_{2}}\right],
\end{equation}
where $\mathcal{O}$ denote differential operators containing covariant derivatives and the Minkowski metric $\eta_{\mu\nu}$, so also the contractions of the Riemann and contorsion tensors are considered in the action. Moreover, the tilde $\,\widetilde{\,}\,$ represents the quantities calculated with respect to the total connection $\widetilde{\Gamma}$. We will expand the previous expression to obtain the general form for the gravitational Lagrangian\footnote{Note that we have more terms than in the Lagrangian written in our previous work \cite{delaCruz-Dombriz:2018aal}. We have added them for completion, although they are completly redundant in the linear regime.}
{\scriptsize
\begin{eqnarray}
\label{lagrangian}
\mathcal{L}_{q}&=&\widetilde{R}\widetilde{F}_{1}\left(\Box\right)\widetilde{R}+\widetilde{R}\widetilde{F}_{2}\left(\Box\right)\partial_{\mu}\partial_{\nu}\widetilde{R}^{\mu\nu}+\widetilde{R}_{\mu\nu}\widetilde{F}_{3}\left(\Box\right)\widetilde{R}^{\left(\mu\nu\right)}+\widetilde{R}_{\mu\nu}\widetilde{F}_{4}\left(\Box\right)\widetilde{R}^{\left[\mu\nu\right]}+\widetilde{R}_{\left(\mu\right.}^{\,\,\,\left.\nu\right)}\widetilde{F}_{5}\left(\Box\right)\partial_{\nu}\partial_{\lambda}\widetilde{R}^{\mu\lambda}+\widetilde{R}_{\left[\mu\right.}^{\,\,\,\left.\nu\right]}\widetilde{F}_{6}\left(\Box\right)\partial_{\nu}\partial_{\lambda}\widetilde{R}^{\mu\lambda}
\nonumber
\\
&+&\widetilde{R}_{\mu}^{\,\,\,\nu}\widetilde{F}_{7}\left(\Box\right)\partial_{\nu}\partial_{\lambda}\widetilde{R}^{\left(\mu\lambda\right)}+\widetilde{R}_{\mu}^{\,\,\,\nu}\widetilde{F}_{8}\left(\Box\right)\partial_{\nu}\partial_{\lambda}\widetilde{R}^{\left[\mu\lambda\right]}+\widetilde{R}^{\lambda\sigma}\widetilde{F}_{9}\left(\Box\right)\partial_{\mu}\partial_{\sigma}\partial_{\nu}\partial_{\lambda}\widetilde{R}^{\mu\nu}+\widetilde{R}_{\left(\mu\lambda\right)}\widetilde{F}_{10}\left(\Box\right)\partial_{\nu}\partial_{\sigma}\widetilde{R}^{\mu\nu\lambda\sigma}+\widetilde{R}_{\left[\mu\lambda\right]}\widetilde{F}_{11}\left(\Box\right)\partial_{\nu}\partial_{\sigma}\widetilde{R}^{\mu\nu\lambda\sigma}
\nonumber
\\
&+&\widetilde{R}_{\mu\lambda}\widetilde{F}_{12}\left(\Box\right)\partial_{\nu}\partial_{\sigma}\widetilde{R}^{\left(\mu\nu\right|\left.\lambda\sigma\right)}+\widetilde{R}_{\mu\lambda}\widetilde{F}_{13}\left(\Box\right)\partial_{\nu}\partial_{\sigma}\widetilde{R}^{\left[\mu\nu\right|\left.\lambda\sigma\right]}+\widetilde{R}_{\mu\nu\lambda\sigma}\widetilde{F}_{14}\left(\Box\right)\widetilde{R}^{\left(\mu\nu\right|\left.\lambda\sigma\right)}+\widetilde{R}_{\mu\nu\lambda\sigma}\widetilde{F}_{15}\left(\Box\right)\widetilde{R}^{\left[\mu\nu\right|\left.\lambda\sigma\right]}+\widetilde{R}_{\left(\rho\mu\right|\left.\nu\lambda\right)}\widetilde{F}_{16}\left(\Box\right)\partial^{\rho}\partial_{\sigma}\widetilde{R}^{\mu\nu\lambda\sigma}
\nonumber
\\
&+&\widetilde{R}_{\left[\rho\mu\right|\left.\nu\lambda\right]}\widetilde{F}_{17}\left(\Box\right)\partial^{\rho}\partial_{\sigma}\widetilde{R}^{\mu\nu\lambda\sigma}+\widetilde{R}_{\rho\mu\nu\lambda}\widetilde{F}_{18}\left(\Box\right)\partial^{\rho}\partial_{\sigma}\widetilde{R}^{\left(\mu\nu\right|\left.\lambda\sigma\right)}+\widetilde{R}_{\rho\mu\nu\lambda}\widetilde{F}_{19}\left(\Box\right)\partial^{\rho}\partial_{\sigma}\widetilde{R}^{\left[\mu\nu\right|\left.\lambda\sigma\right]}+\widetilde{R}_{\left(\mu\nu\right|\left.\rho\sigma\right)}\widetilde{F}_{20}\left(\Box\right)\partial^{\nu}\partial^{\sigma}\partial_{\alpha}\partial_{\beta}\widetilde{R}^{\mu\alpha\rho\beta}
\nonumber
\\
&+&\widetilde{R}_{\left[\mu\nu\right|\left.\rho\sigma\right]}\widetilde{F}_{21}\left(\Box\right)\partial^{\nu}\partial^{\sigma}\partial_{\alpha}\partial_{\beta}\widetilde{R}^{\mu\alpha\rho\beta}+\widetilde{R}_{\mu\nu\rho\sigma}\widetilde{F}_{22}\left(\Box\right)\partial^{\nu}\partial^{\sigma}\partial_{\alpha}\partial_{\beta}\widetilde{R}^{\left(\mu\alpha\right|\left.\rho\beta\right)}+\widetilde{R}_{\mu\nu\rho\sigma}\widetilde{F}_{23}\left(\Box\right)\partial^{\nu}\partial^{\sigma}\partial_{\alpha}\partial_{\beta}\widetilde{R}^{\left[\mu\alpha\right|\left.\rho\beta\right]}+K_{\mu\nu\rho}\widetilde{F}_{24}\left(\Box\right)K^{\mu\nu\rho}
\nonumber
\\
&+&K_{\mu\nu\rho}\widetilde{F}_{25}\left(\Box\right)K^{\mu\rho\nu}+K_{\mu\,\,\rho}^{\,\,\rho}\widetilde{F}_{26}\left(\Box\right)K_{\,\,\,\,\,\sigma}^{\mu\sigma}+K_{\,\,\nu\rho}^{\mu}\widetilde{F}_{27}\left(\Box\right)\partial_{\mu}\partial_{\sigma}K^{\sigma\nu\rho}+K_{\,\,\nu\rho}^{\mu}\widetilde{F}_{28}\left(\Box\right)\partial_{\mu}\partial_{\sigma}K^{\sigma\rho\nu}+K_{\mu\,\,\,\,\,\nu}^{\,\,\rho}\widetilde{F}_{29}\left(\Box\right)\partial_{\rho}\partial_{\sigma}K^{\mu\nu\sigma}
\nonumber
\\
&+&K_{\mu\,\,\,\,\,\nu}^{\,\,\rho}\widetilde{F}_{30}\left(\Box\right)\partial_{\rho}\partial_{\sigma}K^{\mu\sigma\nu}+K_{\,\,\,\,\,\rho}^{\mu\rho}\widetilde{F}_{31}\left(\Box\right)\partial_{\mu}\partial_{\nu}K_{\,\,\,\,\,\sigma}^{\nu\sigma}+K_{\mu}^{\,\,\nu\rho}\widetilde{F}_{32}\left(\Box\right)\partial_{\nu}\partial_{\rho}\partial_{\alpha}\partial_{\sigma}K^{\mu\alpha\sigma}+K_{\,\,\,\lambda\sigma}^{\lambda}\widetilde{F}_{33}\left(\Box\right)\partial_{\rho}\partial_{\nu}K^{\nu\rho\sigma}+\widetilde{R}_{\,\,\nu\rho\sigma}^{\mu}\widetilde{F}_{34}\left(\Box\right)\partial_{\mu}K^{\nu\rho\sigma}
\nonumber
\\
&+&\widetilde{R}_{\mu\nu\,\,\sigma}^{\,\,\,\,\,\,\rho}\widetilde{F}_{35}\left(\Box\right)\partial_{\rho}K^{\mu\nu\sigma}+\widetilde{R}_{\left(\rho\sigma\right)}\widetilde{F}_{36}\left(\Box\right)\partial_{\nu}K^{\nu\rho\sigma}+\widetilde{R}_{\left[\rho\sigma\right]}\widetilde{F}_{37}\left(\Box\right)\partial_{\nu}K^{\nu\rho\sigma}+\widetilde{R}_{\rho\sigma}\widetilde{F}_{38}\left(\Box\right)\partial_{\nu}K^{\rho\nu\sigma}+\widetilde{R}_{\left(\rho\sigma\right)}\widetilde{F}_{39}\left(\Box\right)\partial^{\sigma}K_{\,\,\,\,\,\mu}^{\rho\mu}
\nonumber
\\
&+&\widetilde{R}_{\left[\rho\sigma\right]}\widetilde{F}_{40}\left(\Box\right)\partial^{\sigma}K_{\,\,\,\,\,\mu}^{\rho\mu}+\widetilde{R}\widetilde{F}_{41}\left(\Box\right)\partial_{\rho}K_{\,\,\,\,\,\mu}^{\rho\mu}+\widetilde{R}_{\,\,\alpha\,\,\sigma}^{\mu\,\,\rho}\widetilde{F}_{42}\left(\Box\right)\partial_{\mu}\partial_{\rho}\partial_{\nu}K^{\nu\left(\alpha\sigma\right)}+\widetilde{R}_{\,\,\alpha\,\,\sigma}^{\mu\,\,\rho}\widetilde{F}_{43}\left(\Box\right)\partial_{\mu}\partial_{\rho}\partial_{\nu}K^{\nu\left[\alpha\sigma\right]}+\widetilde{R}_{\,\,\alpha\,\,\sigma}^{\mu\,\,\rho}\widetilde{F}_{44}\left(\Box\right)\partial_{\mu}\partial_{\rho}\partial_{\nu}K^{\alpha\nu\sigma}
\nonumber
\\
&+&\widetilde{R}_{\,\,\left.\sigma\right)}^{\left(\mu\right.}\widetilde{F}_{45}\left(\Box\right)\partial_{\mu}\partial_{\nu}\partial_{\alpha}K^{\sigma\nu\alpha}
+\widetilde{R}_{\,\,\left.\sigma\right]}^{\left[\mu\right.}\widetilde{F}_{46}\left(\Box\right)\partial_{\mu}\partial_{\nu}\partial_{\alpha}K^{\sigma\nu\alpha}+\widetilde{R}_{\mu\nu\lambda\sigma}\widetilde{F}_{47}\left(\Box\right)\widetilde{R}^{\mu\lambda\nu\sigma},
\end{eqnarray}
}
where the $\widetilde{F}_{i}\left(\Box\right)$'s are functions of the D'Alambertian $\Box=\eta_{\mu\nu}\partial^{\mu}\partial^{\nu}$, of the form
\begin{equation}
\label{infinite}
\widetilde{F}_{i}\left(\Box\right)=\stackrel[n=0]{N}{\sum}\widetilde{f}_{i,n}\left(\frac{\Box}{M_{S}}\right)^{n},
\end{equation}
where $M_{S}$ holds for the mass defining the scale at which non-localities starts to play a role. Also, in the previous expression $n$ can be a finite (finite higher-order derivatives theories), or infinite (IDG) number, as we will consider from now onwards unless stated otherwise. However, finite derivatives will incur ghosts and other instabilities. In the last section we shall show how only considering an infinite number of derivatives in Eq.\eqref{infinite} one can avoid the ghosts appearance for the torsion sector, which extends the current results on the metric one \cite{Biswas:2011ar}.

Since one needs to recover the purely metric IDG action when the torsion is zero, there are some constraints in the form of the $\widetilde{F}$ functions. In order to obtain these relations, let us write the action of the metric theory around a Minkowski background as presented in \cite{Biswas:2011ar}
\begin{eqnarray}
\label{metricidg}
&&\mathcal{L}_{\rm IDG}=R{F}_{1}\left(\Box\right)R+R{F}_{2}\left(\Box\right)\partial_{\mu}\partial_{\nu}R^{\mu\nu}+R_{\mu\nu}{F}_{3}\left(\Box\right)R^{\mu\nu}+R_{\mu}^{\,\,\,\nu}{F}_{4}\left(\Box\right)\partial_{\nu}\partial_{\lambda}R^{\mu\lambda}
\nonumber
\\
&+&R^{\lambda\sigma}F_{5}\left(\Box\right)\partial_{\mu}\partial_{\sigma}\partial_{\nu}\partial_{\lambda}R^{\mu\nu}+R_{\mu\lambda}F_{6}\left(\Box\right)\partial_{\nu}\partial_{\sigma}R^{\mu\nu\lambda\sigma}+{R}_{\mu\nu\lambda\sigma}{F}_{7}\left(\Box\right){R}^{\mu\nu\lambda\sigma}
\nonumber
\\
&+&{R}_{\rho\mu\nu\lambda}F_{8}\left(\Box\right)\partial^{\rho}\partial_{\sigma}{R}^{\mu\nu\lambda\sigma}+{R}_{\mu\nu\rho\sigma}F_{9}\left(\Box\right)\partial^{\nu}\partial^{\sigma}\partial_{\alpha}\partial_{\beta}{R}^{\mu\alpha\rho\beta},
\end{eqnarray}
and compare it with the Lagrangian in Eq. \eqref{lagrangian} in the limit when torsion goes to zero
\begin{eqnarray}
\label{nulltorsion}
&&\mathcal{L}_{q}\left(K_{\,\,\,\nu\sigma}^{\mu}\rightarrow0\right)=R\widetilde{F}_{1}\left(\Box\right)R+R\widetilde{F}_{2}\left(\Box\right)\partial_{\mu}\partial_{\nu}R^{\mu\nu}+R_{\mu\nu}\widetilde{F}_{3}\left(\Box\right)R^{\mu\nu}+R_{\mu}^{\,\,\,\nu}\left(\widetilde{F}_{5}\left(\Box\right)+\widetilde{F}_{7}\left(\Box\right)\right)\partial_{\nu}\partial_{\lambda}R^{\mu\lambda}
\nonumber
\\
&+&R^{\lambda\sigma}\widetilde{F}_{9}\left(\Box\right)\partial_{\mu}\partial_{\sigma}\partial_{\nu}\partial_{\lambda}R^{\mu\nu}+R_{\mu\lambda}\left(\widetilde{F}_{10}\left(\Box\right)+\widetilde{F}_{12}\left(\Box\right)\right)\partial_{\nu}\partial_{\sigma}R^{\mu\nu\lambda\sigma}+{R}_{\mu\nu\lambda\sigma}\left(\widetilde{F}_{14}\left(\Box\right)+\frac{\widetilde{F}_{47}\left(\Box\right)}{2}\right){R}^{\mu\nu\lambda\sigma}
\nonumber
\\
&+&{R}_{\rho\mu\nu\lambda}\left(\widetilde{F}_{16}\left(\Box\right)+\widetilde{F}_{18}\left(\Box\right)\right)\partial^{\rho}\partial_{\sigma}{R}^{\mu\nu\lambda\sigma}+{R}_{\mu\nu\rho\sigma}\left(\widetilde{F}_{20}\left(\Box\right)+\widetilde{F}_{22}\left(\Box\right)\right)\partial^{\nu}\partial^{\sigma}\partial_{\alpha}\partial_{\beta}{R}^{\mu\alpha\rho\beta}.
\end{eqnarray}
Then a straightforward comparison between Eqs.\eqref{metricidg} and \eqref{nulltorsion} make it clear that the following relations need to hold
\begin{eqnarray}
&&\tilde{F}_{1}\left(\Box\right)=F_{1}\left(\Box\right),\;\tilde{F}_{2}\left(\Box\right)=F_{2}\left(\Box\right),\;\tilde{F}_{3}\left(\Box\right)=F_{3}\left(\Box\right),\;\tilde{F}_{5}\left(\Box\right)+\tilde{F}_{7}\left(\Box\right)=F_{4}\left(\Box\right),\;\tilde{F}_{9}\left(\Box\right)=F_{5}\left(\Box\right),
\nonumber
\\
&&\,
\\
&&\tilde{F}_{10}\left(\Box\right)+\tilde{F}_{12}\left(\Box\right)=F_{6}\left(\Box\right),\;\tilde{F}_{14}\left(\Box\right)+\frac{\widetilde{F}_{47}\left(\Box\right)}{2}=F_{7}\left(\Box\right),\;\tilde{F}_{16}\left(\Box\right)+\tilde{F}_{18}\left(\Box\right)=F_{8}\left(\Box\right),\;\tilde{F}_{20}\left(\Box\right)+\tilde{F}_{22}\left(\Box\right)=F_{9}\left(\Box\right).\nonumber
\end{eqnarray}
In order to check which are the terms that are of order ${\cal O} (h^{2})$ in the Lagrangian Eq.\eqref{lagrangian}, we still need to substitute the linearized expressions of the curvature tensors
\begin{equation}
\label{riemann}
\tilde{R}_{\mu\nu\rho\lambda}=\partial_{\left[\nu\right.}\partial_{\rho}h_{\left.\lambda\mu\right]}-\partial_{\left[\nu\right.}\partial_{\lambda}h_{\left.\rho\mu\right]}+2\partial_{\left[\nu\right.}K_{\rho\left|\mu\right]\lambda},
\end{equation}
\begin{equation}
\label{ricci}
\tilde{R}_{\mu\nu}=\partial_{\sigma}\partial_{\left(\nu\right.}h_{\left.\mu\right)}^{\,\,\,\sigma}-\frac{1}{2}\left(\partial_{\mu}\partial_{\nu}h+\Box h_{\mu\nu}\right)-\partial_{\sigma}K_{\,\,\,\mu\nu}^{\sigma}+\partial_{\mu}K_{\,\,\,\sigma\nu}^{\sigma},
\end{equation}
\begin{equation}
\label{scalar}
\tilde{R}=\partial_{\mu}\partial_{\nu}h^{\mu\nu}-\Box h-2\partial_{\mu}K_{\,\,\,\,\,\,\,\nu}^{\mu\nu},
\end{equation}
where $\left(\mu\nu\right)$ and $\left[\mu\nu\right]$ represent the symmetrisation and anti-symmetrisation of indices respectively. We have computed each term appearing in the Lagrangian Eq.\eqref{lagrangian} separately. These explicit calculations can be found in Appendix \ref{ap1}. Finally, using the above expressions obtained and further simplification would yield  the linearized action for metric, torsion and the mixed terms:
\begin{equation}
\label{laaccion}
S=-\int {\rm d}^{4}x\left(\mathcal{L}_{M}+\mathcal{L}_{MT}+\mathcal{L}_{T}\right)=S_{M}+S_{MT}+S_{T},
\end{equation}
where
\begin{equation}
\label{compac1}
\mathcal{L}_{M}=\frac{1}{2}h_{\mu\nu}\Box a\left(\Box\right)h^{\mu\nu}+h_{\mu}^{\,\,\alpha}b\left(\Box\right)\partial_{\alpha}\partial_{\sigma}h^{\sigma\mu}+hc\left(\Box\right)\partial_{\mu}\partial_{\nu}h^{\mu\nu}+\frac{1}{2}h\Box d\left(\Box\right)h+h^{\lambda\sigma}\frac{f\left(\Box\right)}{\Box}\partial_{\sigma}\partial_{\lambda}\partial_{\mu}\partial_{\nu}h^{\mu\nu},
\end{equation}
\begin{equation}
\label{compac2}
\mathcal{L}_{MT}=h\Box u\left(\Box\right)\partial_{\rho}K_{\,\,\,\,\,\sigma}^{\rho\sigma}+h_{\mu\nu}v_{1}\left(\Box\right)\partial^{\mu}\partial^{\nu}\partial_{\rho}K_{\,\,\,\,\,\sigma}^{\rho\sigma}+h_{\mu\nu}v_{2}\left(\Box\right)\partial^{\nu}\partial_{\sigma}\partial_{\rho}K^{\mu\sigma\rho}+h_{\mu\nu}\Box w\left(\Box\right)\partial_{\rho}K^{\rho\mu\nu},
\end{equation}
\begin{eqnarray}
\label{compac3}
\mathcal{L}_{T}&=&K^{\mu\sigma\lambda}p_{1}\left(\Box\right)K_{\mu\sigma\lambda}+K^{\mu\sigma\lambda}p_{2}\left(\Box\right)K_{\mu\lambda\sigma}+K_{\mu\,\,\rho}^{\,\,\rho}p_{3}\left(\Box\right)K_{\,\,\,\,\,\sigma}^{\mu\sigma}+K_{\,\,\nu\rho}^{\mu}q_{1}\left(\Box\right)\partial_{\mu}\partial_{\sigma}K^{\sigma\nu\rho}+K_{\,\,\nu\rho}^{\mu}q_{2}\left(\Box\right)\partial_{\mu}\partial_{\sigma}K^{\sigma\rho\nu}
\nonumber
\\
&+&K_{\mu\,\,\,\,\,\nu}^{\,\,\rho}q_{3}\left(\Box\right)\partial_{\rho}\partial_{\sigma}K^{\mu\nu\sigma}+K_{\mu\,\,\,\,\,\nu}^{\,\,\rho}q_{4}\left(\Box\right)\partial_{\rho}\partial_{\sigma}K^{\mu\sigma\nu}+K_{\,\,\,\,\,\rho}^{\mu\rho}q_{5}\left(\Box\right)\partial_{\mu}\partial_{\nu}K_{\,\,\,\,\,\sigma}^{\nu\sigma}+K_{\,\,\,\lambda\sigma}^{\lambda}q_{6}\left(\Box\right)\partial_{\mu}\partial_{\alpha}K^{\sigma\mu\alpha}
\nonumber
\\
&+&K_{\mu}^{\,\,\nu\rho}s\left(\Box\right)\partial_{\nu}\partial_{\rho}\partial_{\alpha}\partial_{\sigma}K^{\mu\alpha\sigma}.
\end{eqnarray}
In order to get a deeper insight about how the functions involved in Eqs.\eqref{compac1}, \eqref{compac2} and \eqref{compac3} are related with the $\tilde{F}_{i}\left(\Box\right)$'s in Eq.\eqref{lagrangian}, we refer the readers to Appendix \ref{ap2}.
At this stage, it is interesting to note that $\mathcal{L}_{M}$ in Eq.\eqref{compac1} possesses metric terms only and coincides with the Lagrangian of the non-torsion case~\cite{Biswas:2011ar}, as expected. On the other hand, $\mathcal{L}_{MT}$ in Eq.\eqref{compac2} contains the mixed terms between metric and torsion, whereas $\mathcal{L}_{T}$ contains only torsion expressions.  It is also worth calculating the local limit by taking $M_{S} \rightarrow \infty$. For the detailed calculations we refer the reader to Appendix \ref{ap3}. Here we will just summarise that the local limit of the theory is 
\begin{eqnarray}
\mathcal{L}_{{\rm GPG}}&=&\tilde{R}+b_{1}\tilde{R}^{2}+b_{2}\tilde{R}_{\mu\nu\rho\sigma}\tilde{R}^{\mu\nu\rho\sigma}+b_{3}\tilde{R}_{\mu\nu\rho\sigma}\tilde{R}^{\rho\sigma\mu\nu}+2\left(b_{1}-b_{2}-b_{3}\right)\tilde{R}_{\mu\nu\rho\sigma}\tilde{R}^{\mu\rho\nu\sigma}+b_{5}\tilde{R}_{\mu\nu}\tilde{R}^{\mu\nu}
\nonumber
\\
&&-\left(4b_{1}+b_{5}\right)\tilde{R}_{\mu\nu}\tilde{R}^{\nu\mu}+a_{1}K_{\mu\nu\rho}K^{\mu\nu\rho}+a_{2}K_{\mu\nu\rho}K^{\mu\rho\nu}+a_{3}K_{\nu\,\,\,\,\,\mu}^{\,\,\,\mu}K_{\,\,\,\,\,\,\rho}^{\nu\rho}+c_{1}K_{\,\,\nu\rho}^{\mu}\nabla_{\mu}\nabla_{\sigma}K^{\sigma\nu\rho}
\nonumber
\\
&&+c_{2}K_{\,\,\nu\rho}^{\mu}\nabla_{\mu}\nabla_{\sigma}K^{\sigma\rho\nu}+c_{3}K_{\mu\,\,\,\,\,\nu}^{\,\,\rho}\nabla_{\rho}\nabla_{\sigma}K^{\mu\nu\sigma}+c_{4}K_{\mu\,\,\,\,\,\nu}^{\,\,\rho}\nabla_{\rho}\nabla_{\sigma}K^{\mu\sigma\nu},
\end{eqnarray} 
given that the conditions in \eqref{localconditions} meet.

The fact that the terms of the form $\nabla_{\mu}K_{\,\,\,\nu\rho}^{\mu}\nabla_{\sigma}K^{\sigma\nu\rho}$ are part of the Lagrangian has been proven recently to be a sufficient condition to make the vector modes present in the theory ghost-free in the IR limit \cite{Jimenez:2019qjc}.

In our previous work \cite{delaCruz-Dombriz:2018aal} the authors showed that one of these modes can be ghost-free also in the UV limit. In this work we shall prove that the two modes can be made ghost-free in this limit.

\section{Field equations}
\label{sec:III}

Since the connection under consideration is different from Levi-Civita one, and consequently the metric and the connections are {\it a priori} independent, we need to apply Palatini formalism to obtain the field equations.   We will have two set of equations;
\begin{itemize}
\item \textbf{Einstein Equations}: Variation of the action, Eq.\eqref{laaccion}, with respect to the metric:
\begin{equation}
\frac{\delta_{g}S_{M}}{\delta g^{\mu\nu}}+\frac{\delta_{g}S_{MT}}{\delta g^{\mu\nu}}=0.
\end{equation}
\item \textbf{Cartan Equations}: Variation of the action, Eq.\eqref{laaccion}, with respect to the contorsion
\begin{equation}
\frac{\delta_{K}S_{MT}}{\delta K_{\,\,\nu\rho}^{\mu}}+\frac{\delta_{K}S_{T}}{\delta K_{\,\,\nu\rho}^{\mu}}=0.
\end{equation}
\end{itemize}
It is interesting to note that $\frac{\delta_{g}S_{M}}{\delta g^{\mu\nu}}$ has already been calculated in Ref.~\cite{Biswas:2011ar}, although, calculations involving such a term were performed again as a consistency check. Let us sketch the calculations leading towards the field equations.

\subsection{Einstein Equations}

Performing variations with respect to the metric in $S_{M}$, we find
\begin{equation}
\frac{\delta_{g}S_{M}}{\delta g^{\mu\nu}}=\Box a\left(\Box\right)h_{\mu\nu}+b\left(\Box\right)\partial_{\sigma}\partial_{\left(\nu\right.}h_{\left.\mu\right)}^{\,\,\,\sigma}+c\left(\Box\right)\left[\partial_{\mu}\partial_{\nu}h+\eta_{\mu\nu}\partial_{\rho}\partial_{\sigma}h^{\rho\sigma}\right]+\eta_{\mu\nu}\Box d\left(\Box\right)h+2\frac{f\left(\Box\right)}{\Box}\partial_{\mu}\partial_{\nu}\partial_{\rho}\partial_{\sigma}h^{\rho\sigma},
\end{equation}
which is compatible with the results in Ref.~\cite{Biswas:2011ar}.
For $S_{MT}$, we have
\begin{equation}
\frac{\delta_{g}S_{MT}}{\delta g^{\mu\nu}}=\eta_{\mu\nu}\Box u\left(\Box\right)\partial_{\rho}K_{\,\,\,\,\,\sigma}^{\rho\sigma}+v_{1}\left(\Box\right)\partial_{\mu}\partial_{\nu}\partial_{\rho}K_{\,\,\,\,\,\sigma}^{\rho\sigma}+v_{2}\left(\Box\right)\partial_{\sigma}\partial_{\rho}\partial_{\left(\nu\right.}K_{\left.\mu\right)}^{\,\,\,\sigma\rho}+\Box w\left(\Box\right)\partial_{\rho}K_{\,\,\left(\mu\nu\right)}^{\rho}.
\end{equation}
Therefore, the resulting Einstein's equations are:
\begin{eqnarray}
\label{einsteingen}
&&\Box a\left(\Box\right)h_{\mu\nu}+b\left(\Box\right)\partial_{\sigma}\partial_{\left(\nu\right.}h_{\left.\mu\right)}^{\,\,\,\sigma}+c\left(\Box\right)\left[\partial_{\mu}\partial_{\nu}h+\eta_{\mu\nu}\partial_{\rho}\partial_{\sigma}h^{\rho\sigma}\right]+\eta_{\mu\nu}\Box d\left(\Box\right)h+2\frac{f\left(\Box\right)}{\Box}\partial_{\mu}\partial_{\nu}\partial_{\rho}\partial_{\sigma}h^{\rho\sigma}
\nonumber
\\
&&+\eta_{\mu\nu}\Box u\left(\Box\right)\partial_{\rho}K_{\,\,\,\,\,\sigma}^{\rho\sigma}+v_{1}\left(\Box\right)\partial_{\mu}\partial_{\nu}\partial_{\rho}K_{\,\,\,\,\,\sigma}^{\rho\sigma}+v_{2}\left(\Box\right)\partial_{\sigma}\partial_{\rho}\partial_{\left(\nu\right.}K_{\left.\mu\right)}^{\,\,\,\sigma\rho}+\Box w\left(\Box\right)\partial_{\rho}K_{\,\,\left(\mu\nu\right)}^{\rho}=\tau_{\mu\nu},
\end{eqnarray}
where $\tau_{\mu\nu}={\delta S_{matter}}/{\delta g^{\mu\nu}}$ is the usual energy-momentum tensor for matter fields.
At this stage, we can resort to the conservation of the energy-momentum tensor, $\partial_{\mu}\tau^{\mu\nu}=0$, to find the following constraints on the functions involved in Eq. \eqref{einsteingen}
\begin{eqnarray}
\label{constraints}
&&a(\Box)+b(\Box)=0,~c(\Box)+d(\Box)=0,~b(\Box)+c(\Box)+f(\Box)=0\,,
\nonumber
\\
&&u(\Box)+v_{1}(\Box)=0\,,~~~~~~v_{2}(\Box)-w(\Box)=0\,.
\end{eqnarray}
We can also prove these constraints by looking at the explicit expression of the functions in Eq.\eqref{constraints} in the Appendix \ref{ap2}. 


\subsection{Cartan Equations}

On the other hand, performing variations with respect to the contorsion, we find
\begin{equation}
\label{cartan1}
\frac{\delta S_{MT}}{\delta K_{\,\,\nu\rho}^{\mu}}=-\Box u\left(\Box\right)\partial_{\left[\mu\right.}\eta^{\left.\rho\right]\nu}h-v_{1}\left(\Box\right)\partial^{\alpha}\partial^{\beta}\partial_{\left[\mu\right.}\eta^{\left.\rho\right]\nu}h_{\alpha\beta}-v_{2}\left(\Box\right)\partial^{\beta}\partial^{\nu}\partial^{\left[\rho\right.}h_{\left.\mu\right]\beta}-\Box w\left(\Box\right)\partial_{\left[\mu\right.}h^{\left.\rho\right]\nu},
\end{equation}
and
\begin{eqnarray}
\label{cartan2}
\frac{\delta S_{T}}{\delta K_{\,\,\nu\rho}^{\mu}}&=&2p_{1}\left(\Box\right)K_{\mu}^{\,\,\nu\rho}+2p_{2}\left(\Box\right)K_{\left[\mu\right.}^{\,\,\,\,\,\left.\rho\right]\nu}+2p_{3}\left(\Box\right)\eta^{\nu\left[\rho\right.}K_{\left.\mu\right]\,\,\,\,\,\sigma}^{\,\,\,\,\sigma}-2q_{1}\left(\Box\right)\partial_{\sigma}\partial_{\left[\mu\right.}K^{\left.\rho\right]\nu\sigma}+2q_{2}\left(\Box\right)\partial_{\sigma}\partial_{\left[\mu\right.}K^{\sigma\left|\rho\right]\nu}
\nonumber
\\
&+&q_{3}\left(\Box\right)\left(\partial^{\nu}\partial_{\sigma}K_{\left[\mu\right.}^{\,\,\,\left.\rho\right]\sigma}+\partial_{\sigma}\partial^{\left[\rho\right.}K_{\left.\mu\right]\,\,\,\,\,}^{\,\,\,\,\sigma\nu}\right)+2q_{4}\left(\Box\right)\partial^{\nu}\partial_{\sigma}K_{\mu}^{\,\,\,\sigma\rho}+2q_{5}\left(\Box\right)\eta^{\nu\left[\rho\right.}\partial_{\left.\mu\right]}\partial_{\lambda}K_{\,\,\,\,\,\sigma}^{\lambda\sigma}
\nonumber
\\
&+&q_{6}\left(\Box\right)\left(\partial_{\lambda}\partial_{\alpha}\eta_{\left[\mu\right.}^{\nu}K^{\left.\rho\right]\lambda\alpha}-\partial^{\nu}\partial^{\left[\rho\right.}K_{\left.\mu\right]\lambda}^{\,\,\,\,\,\,\lambda}\right)+2s\left(\Box\right)\partial^{\sigma}\partial^{\lambda}\partial^{\nu}\partial^{\left[\rho\right.}K_{\left.\mu\right]\sigma\lambda}.
\end{eqnarray}
This leads us to the Cartan Equations;
\begin{eqnarray}
\label{cartangen}
&-&\Box u\left(\Box\right)\partial_{\left[\mu\right.}\eta^{\left.\rho\right]\nu}h-v_{1}\left(\Box\right)\partial^{\alpha}\partial^{\beta}\partial_{\left[\mu\right.}\eta^{\left.\rho\right]\nu}h_{\alpha\beta}-v_{2}\left(\Box\right)\partial^{\beta}\partial^{\nu}\partial^{\left[\rho\right.}h_{\left.\mu\right]\beta}-\Box w\left(\Box\right)\partial_{\left[\mu\right.}h^{\left.\rho\right]\nu}+2p_{1}\left(\Box\right)K_{\mu}^{\,\,\nu\rho}+2p_{2}\left(\Box\right)K_{\left[\mu\right.}^{\,\,\,\,\,\left.\rho\right]\nu}
\nonumber
\\
&+&2p_{3}\left(\Box\right)\eta^{\nu\left[\rho\right.}K_{\left.\mu\right]\,\,\,\,\,\sigma}^{\,\,\,\,\sigma}-2q_{1}\left(\Box\right)\partial_{\sigma}\partial_{\left[\mu\right.}K^{\left.\rho\right]\nu\sigma}+2q_{2}\left(\Box\right)\partial_{\sigma}\partial_{\left[\mu\right.}K^{\sigma\left|\rho\right]\nu}+q_{3}\left(\Box\right)\left(\partial^{\nu}\partial_{\sigma}K_{\left[\mu\right.}^{\,\,\,\left.\rho\right]\sigma}+\partial_{\sigma}\partial^{\left[\rho\right.}K_{\left.\mu\right]\,\,\,\,\,}^{\,\,\,\,\sigma\nu}\right)
\nonumber
\\
&+&2q_{4}\left(\Box\right)\partial^{\nu}\partial_{\sigma}K_{\mu}^{\,\,\,\sigma\rho}+2q_{5}\left(\Box\right)\eta^{\nu\left[\rho\right.}\partial_{\left.\mu\right]}\partial_{\lambda}K_{\,\,\,\,\,\sigma}^{\lambda\sigma}+q_{6}\left(\Box\right)\left(\partial_{\lambda}\partial_{\alpha}\eta_{\left[\mu\right.}^{\nu}K^{\left.\rho\right]\lambda\alpha}-\partial^{\nu}\partial^{\left[\rho\right.}K_{\left.\mu\right]\lambda}^{\,\,\,\,\,\,\lambda}\right)
\nonumber
\\
&+&2s\left(\Box\right)\partial^{\sigma}\partial^{\lambda}\partial^{\nu}\partial^{\left[\rho\right.}K_{\left.\mu\right]\sigma\lambda}=\Sigma_{\mu}^{\,\,\,\nu\rho},
\end{eqnarray}
where $\Sigma_{\mu}^{\,\,\,\nu\rho}={\delta S_{matter}}/{\delta K_{\,\,\,\nu\rho}^{\mu}}$. From these field equations Eqs. \eqref{einsteingen} and \eqref{cartangen} exact solutions cannot be obtained so easily. In order to solve them, we will decompose the contorsion field $K_{\mu\nu\rho}$ into its three irreducible components.

\section{Torsion decomposition}
\label{sec:IV}

In four dimensions, the torsion field $T_{\mu\nu\rho}$, as well as the contorsion field $K_{\mu\nu\rho}$, can be decomposed into three irreducible Lorentz invariant terms \cite{Shapiro:2001rz}, yielding
\begin{equation}
\label{decom}
\begin{cases}
\textrm{Trace contorsion vector:}\,\,T_{\mu}=K_{\mu\,\,\,\,\,\nu}^{\,\,\,\nu},\\
\,\\
\textrm{Axial contorsion vector:}\,\,S^{\mu}=\varepsilon^{\rho\sigma\nu\mu}K_{\rho\sigma\nu},\\
\,\\
\textrm{Tensor}\,\,q_{\,\,\nu\rho}^{\mu},\,\,{\rm such\,that}\;\;q_{\,\,\mu\nu}^{\nu}=0\,\,\textrm{and}\,\,\varepsilon^{\rho\sigma\nu\mu}q_{\rho\sigma\nu}=0,
\end{cases}
\end{equation}
such that the contorsion field becomes
\begin{equation}
\label{decom2}
K_{\mu\nu\rho}=\frac{1}{3}\left(T_{\mu}g_{\nu\rho}-T_{\rho}g_{\nu\mu}\right)-\frac{1}{6}\varepsilon_{\mu\nu\rho\sigma}S^{\sigma}+q_{\mu\nu\rho}.
\end{equation}
This decomposition turns out to be very useful, thanks to the fact that the three terms in Eq.\eqref{decom} propagate different dynamical off-shell degrees of freedom. Hence, it is better to study them separately, compared to all the torsion contribution at the same time. Also interaction with matter, more specifically with fermions, is only made via the axial vector \cite{Shapiro:2001rz}. That is why the two remaining components are usually known as {\it inert torsion}. Under this decomposition we will study how the torsion related terms in the linearised Lagrangian in Eq.\eqref{laaccion} change, and how to derive the corresponding field equations.

Introducing \eqref{decom2}, and the constrains of the functions in \eqref{constraints}, in \eqref{compac2} we find that the mixed term of the Lagrangian becomes
\begin{eqnarray}
\label{mixedecom}
\mathcal{L}_{MT}&=&h\Box\left(u\left(\Box\right)+\frac{1}{3}v_{2}\left(\Box\right)\right)\partial_{\mu}T^{\mu}-h_{\mu\nu}\left(u\left(\Box\right)+\frac{1}{3}v_{2}\left(\Box\right)\right)\partial^{\mu}\partial^{\nu}\partial_{\rho}T^{\rho}+h_{\mu\nu}v_{2}\left(\Box\right)\partial^{\nu}\partial_{\rho}\partial_{\sigma}q^{\mu\rho\sigma}
\nonumber
\\
&+&h_{\mu\nu}\Box v_{2}\left(\Box\right)\partial_{\sigma}q^{\mu\nu\sigma}.
\end{eqnarray}
Now, integrating by parts and using the linearised expression for the Ricci scalar we find
\begin{equation}
\label{mixedecom2}
\mathcal{L}_{MT}=-R\left(u\left(\Box\right)+\frac{1}{3}v_{2}\left(\Box\right)\right)\partial_{\mu}T^{\mu}+h_{\mu\nu}v_{2}\left(\Box\right)\partial^{\nu}\partial_{\rho}\partial_{\sigma}q^{\mu\rho\sigma}+h_{\mu\nu}\Box v_{2}\left(\Box\right)\partial_{\sigma}q^{\mu\nu\sigma}
\end{equation}
The first term accounts for a non-minimal coupling of the trace vector with the curvature, that is known for producing ghostly degrees of freedom \cite{Jimenez:2008sq}. Therefore, for stability reasons we impose $v_{2}\left(\Box\right)=-3u\left(\Box\right)$, finally obtaining
\begin{equation}
\label{mixedecom3}
\mathcal{L}_{MT}=-3h_{\mu\nu}u\left(\Box\right)\partial^{\nu}\partial_{\rho}\partial_{\sigma}q^{\mu\rho\sigma}-3h_{\mu\nu}\Box u\left(\Box\right)\partial_{\sigma}q^{\mu\nu\sigma}.
\end{equation}
In order to obtain the pure torsion part of the Lagrangian we substitute \eqref{decom2} into \eqref{compac3}
\begin{eqnarray}
\mathcal{L}_{T}&=&\frac{1}{6}S_{\mu}\left(p_{2}\left(\Box\right)-p_{1}\left(\Box\right)\right)S^{\mu}+\frac{1}{9}\partial_{\left[\mu\right.}S_{\left.\nu\right]}\left(q_{1}\left(\Box\right)-q_{2}\left(\Box\right)-q_{3}\left(\Box\right)+q_{4}\left(\Box\right)\right)\partial^{\left[\mu\right.}S^{\left.\nu\right]}
\nonumber
\\
&+&\frac{1}{3}T_{\mu}\left(2p_{1}\left(\Box\right)+p_{2}\left(\Box\right)+3p_{3}\left(\Box\right)+\frac{1}{2}s\left(\Box\right)\Box^{2}\right)T^{\mu}-\frac{2}{9}\partial_{\left[\mu\right.}T_{\left.\nu\right]}\left(q_{1}\left(\Box\right)+q_{3}\left(\Box\right)+2q_{4}\left(\Box\right)-3q_{6}\left(\Box\right)\right)\partial^{\left[\mu\right.}T^{\left.\nu\right]}
\nonumber
\\
&-&\frac{1}{9}\partial_{\mu}T^{\mu}\left(3q_{1}\left(\Box\right)+3q_{2}\left(\Box\right)+9q_{5}\left(\Box\right)-s\left(\Box\right)\Box\right)\partial_{\nu}T^{\nu}+q_{\mu\nu\rho}p_{1}\left(\Box\right)q^{\mu\nu\rho}+q_{\mu\nu\rho}p_{2}\left(\Box\right)q^{\mu\rho\nu}
\nonumber
\\
&+&q_{\,\,\nu\rho}^{\mu}q_{1}\left(\Box\right)\partial_{\mu}\partial_{\sigma}q^{\sigma\nu\rho}+q_{\,\,\nu\rho}^{\mu}q_{2}\left(\Box\right)\partial_{\mu}\partial_{\sigma}q^{\sigma\rho\nu}+q_{\mu\,\,\,\,\,\nu}^{\,\,\rho}q_{3}\left(\Box\right)\partial_{\rho}\partial_{\sigma}q^{\mu\nu\sigma}+q_{\mu\,\,\,\,\,\nu}^{\,\,\rho}q_{4}\left(\Box\right)\partial_{\rho}\partial_{\sigma}q^{\mu\sigma\nu}
\nonumber
\\
&+&q^{\mu\nu\rho}s\left(\Box\right)\partial_{\nu}\partial_{\rho}\partial_{\sigma}\partial_{\lambda}q_{\mu}^{\,\,\,\sigma\lambda}+\frac{1}{3}T_{\mu}\left(2q_{1}\left(\Box\right)+2q_{3}\left(\Box\right)+4q_{4}\left(\Box\right)-3q_{6}\left(\Box\right)+2s\left(\Box\right)\Box\right)\partial_{\nu}\partial_{\rho}q^{\mu\nu\rho}
\nonumber
\\
&+&\frac{1}{2}\varepsilon_{\mu\nu\rho\sigma}q^{\rho\lambda\sigma}q_{3}\left(\Box\right)\partial_{\lambda}\partial^{\nu}S^{\mu}
\end{eqnarray}
Now we can proceed to calculate the field equations under the torsion decomposition. Varying the Lagrangian with respect to the metric we find the Einstein Equations:
\begin{eqnarray}
\label{einsteindecomposed}
&&\Box a\left(\Box\right)h_{\mu\nu}+b\left(\Box\right)\partial_{\sigma}\partial_{\left(\nu\right.}h_{\left.\mu\right)}^{\,\,\,\sigma}+c\left(\Box\right)\left[\partial_{\mu}\partial_{\nu}h+\eta_{\mu\nu}\partial_{\rho}\partial_{\sigma}h^{\rho\sigma}\right]+\eta_{\mu\nu}\Box d\left(\Box\right)h+2\frac{f\left(\Box\right)}{\Box}\partial_{\mu}\partial_{\nu}\partial_{\rho}\partial_{\sigma}h^{\rho\sigma}
\nonumber
\\
&&-3u\left(\Box\right)\partial_{\sigma}\partial_{\rho}\partial_{\left(\nu\right.}q_{\left.\mu\right)}^{\,\,\,\sigma\rho}-3\Box u\left(\Box\right)\partial_{\rho}q_{\,\,\left(\mu\nu\right)}^{\rho}=\tau_{\mu\nu},
\end{eqnarray}
where we can see that the vectorial parts of the torsion tensor do not contribute.\\
On the other hand, performing variations with respect to the three invariants we find three Cartan Equations
\begin{itemize}
\item Variations with respect to the axial vector $S^{\mu}$
\begin{eqnarray}
\label{cartandecax}
&&\frac{1}{6}\left(p_{2}\left(\Box\right)-p_{1}\left(\Box\right)\right)S_{\mu}+\frac{1}{18}\left(q_{1}\left(\Box\right)-q_{2}\left(\Box\right)-q_{3}\left(\Box\right)+q_{4}\left(\Box\right)\right)\left(\partial_{\mu}\partial_{\nu}S^{\nu}-\Box S_{\mu}\right)
\nonumber
\\
&&+\frac{1}{2}\varepsilon_{\mu\nu\rho\sigma}q_{3}\left(\Box\right)\partial_{\lambda}\partial^{\nu}q^{\rho\lambda\sigma}=\frac{\delta\mathcal{L}_{matter}}{\delta S^{\mu}}.
\end{eqnarray}
\item Variations with respect to the trace vector $T^{\mu}$
\begin{eqnarray}
\label{cartandectra}
&&\frac{1}{3}\left(2p_{1}\left(\Box\right)+p_{2}\left(\Box\right)+3p_{3}\left(\Box\right)+\frac{1}{2}s\left(\Box\right)\Box^{2}\right)T_{\mu}-\frac{1}{9}\left(q_{1}\left(\Box\right)+q_{3}\left(\Box\right)+2q_{4}\left(\Box\right)-3q_{6}\left(\Box\right)\right)\left(\partial_{\mu}\partial_{\nu}T^{\nu}-\Box T_{\mu}\right)
\nonumber
\\
&&+\frac{1}{9}\left(3q_{1}\left(\Box\right)+3q_{2}\left(\Box\right)+9q_{5}\left(\Box\right)-s\left(\Box\right)\Box\right)\partial_{\mu}\partial_{\nu}T^{\nu}
\nonumber
\\
&&+\frac{1}{3}\left(2q_{1}\left(\Box\right)+2q_{3}\left(\Box\right)+4q_{4}\left(\Box\right)-3q_{6}\left(\Box\right)+2s\left(\Box\right)\Box\right)\partial_{\nu}\partial_{\rho}q_{\mu}^{\,\,\nu\rho}=\frac{\delta\mathcal{L}_{matter}}{\delta T^{\mu}}.
\end{eqnarray}
\item Variations with respect to the tensor part $q^{\mu\nu\rho}$
\begin{eqnarray}
&&p_{1}\left(\Box\right)q_{\mu\nu\rho}+p_{2}\left(\Box\right)q_{\left[\mu\rho\right]\nu}+q_{1}\left(\Box\right)\partial_{\left[\mu\right.}\partial_{\sigma}q_{\,\,\nu\left.\rho\right]}^{\sigma}+q_{2}\left(\Box\right)\partial_{\left[\mu\right.}\partial_{\sigma}q_{\,\,\left.\rho\right]\nu}^{\sigma}+q_{3}\left(\Box\right)\partial_{\sigma}\partial_{\left[\rho\right.}q_{\left.\mu\right]\,\,\,\,\,\nu}^{\,\,\,\sigma}+q_{4}\left(\Box\right)\partial_{\nu}\partial_{\sigma}q_{\mu\,\,\,\,\,\rho}^{\,\,\sigma}
\nonumber
\\
&&+s\left(\Box\right)\partial_{\nu}\partial_{\sigma}\partial_{\lambda}\partial_{\left[\rho\right.}q_{\left.\mu\right]}^{\,\,\,\sigma\lambda}+\frac{1}{3}\left(2q_{1}\left(\Box\right)+2q_{3}\left(\Box\right)+4q_{4}\left(\Box\right)-3q_{6}\left(\Box\right)+2s\left(\Box\right)\Box\right)\partial_{\nu}\partial_{\rho}T_{\mu}=\frac{\delta\mathcal{L}_{matter}}{\delta q^{\mu\nu\rho}}.
\end{eqnarray}
\end{itemize}
These decomposed equations will help us to find exact solutions of the theory, as we will see in the following section.

\section{Solutions}
\label{sec:V}

In Ref. \cite{delaCruz-Dombriz:2018aal}, the authors found a particular solution of the IDG with torsion with a fermionic source, where only the axial torsion was considered to be dynamic. Therefore, the Einstein and Cartan equations decoupled and the solutions of the metric were the same as that in the case of IDG. In the following, provided there exists a fermion as a source, and assuming that both axial and trace torsion are different from zero\footnote{The fact that the traceless tensor part of the torsion $q_{\,\,\nu\rho}^{\mu}$ is considered to be negligible is motivated by the fact that in a completely symmetric spacetime this component is identically zero \cite{Sur:2013aia}.}, we will show that we can find additional solutions that were not present in the metric IDG theory. For the IDG theory, solutions were presented in \cite{Buoninfante:2018stt}. In order to make our case more clear, we have divided the calculations in the following two subsections. In the first one, we will solve Cartan equations to obtain the torsion tensor, while in the second one we will solve Einstein equations for the metric tensor.

\subsection{Cartan Equations} 
Let us write down the linearised Lagrangian decomposed into the two vector invariants, where the tensor component of the torsion has been set to zero. Thus,
\begin{eqnarray}
\label{vectoraction}
\mathcal{L}&=&\mathcal{L}_{M}+\frac{1}{6}S_{\mu}\left(p_{2}\left(\Box\right)-p_{1}\left(\Box\right)\right)S^{\mu}+\frac{1}{9}\partial_{\left[\mu\right.}S_{\left.\nu\right]}\left(q_{1}\left(\Box\right)-q_{2}\left(\Box\right)-q_{3}\left(\Box\right)+q_{4}\left(\Box\right)\right)\partial^{\left[\mu\right.}S^{\left.\nu\right]}
\nonumber
\\
&&+\frac{1}{3}T_{\mu}\left(2p_{1}\left(\Box\right)+p_{2}\left(\Box\right)+3p_{3}\left(\Box\right)+\frac{1}{2}s\left(\Box\right)\Box^{2}\right)T^{\mu}-\frac{2}{9}\partial_{\left[\mu\right.}T_{\left.\nu\right]}\left(q_{1}\left(\Box\right)+q_{3}\left(\Box\right)+2q_{4}\left(\Box\right)-3q_{6}\left(\Box\right)\right)\partial^{\left[\mu\right.}T^{\left.\nu\right]}
\nonumber
\\
&&-\frac{1}{9}\partial_{\mu}T^{\mu}\left(3q_{1}\left(\Box\right)+3q_{2}\left(\Box\right)+9q_{5}\left(\Box\right)-s\left(\Box\right)\Box\right)\partial_{\nu}T^{\nu},
\end{eqnarray}
where we have taken into account the constraints on the functions in \eqref{constraints} and the stability condition for the trace vector found in the previous section, namely $v_{2}\left(\Box\right)=-3u\left(\Box\right)$. Due to these conditions, there are no mixed terms between metric and torsion, so the Cartan and Einstein Equations would be decoupled. \\
Despite these constraints, the torsion part of the Lagrangian \eqref{vectoraction} is far from being stable, so before finding some solutions we need to explore under which form of the functions the theory does not have any pathologies.\\
By taking a closer look at \eqref{vectoraction} we realise that, as it is usual in metric IDG, we can make the combinations of the non-local functions to be described by an entire function, which does not introduce any new poles in the propagators, so that we can use the same stability arguments as in the local theory. This means that
\begin{eqnarray}
&&p_{2}\left(\Box\right)-p_{1}\left(\Box\right)=C_{1}e^{-\frac{\Box}{M_{S}^{2}}},
\nonumber
\\
&&q_{1}\left(\Box\right)-q_{2}\left(\Box\right)-q_{3}\left(\Box\right)+q_{4}\left(\Box\right)=C_{2}e^{-\frac{\Box}{M_{S}^{2}}},
\nonumber
\\
&&2p_{1}\left(\Box\right)+p_{2}\left(\Box\right)+3p_{3}\left(\Box\right)+\frac{1}{2}s\left(\Box\right)\Box^{2}=C_{3}e^{-\frac{\Box}{M_{S}^{2}}},
\\
&&q_{1}\left(\Box\right)+q_{3}\left(\Box\right)+2q_{4}\left(\Box\right)-3q_{6}\left(\Box\right)=C_{4}e^{-\frac{\Box}{M_{S}^{2}}},
\nonumber
\\
&&3q_{1}\left(\Box\right)+3q_{2}\left(\Box\right)+9q_{5}\left(\Box\right)-s\left(\Box\right)\Box=C_{5}e^{-\frac{\Box}{M_{S}^{2}}}, \nonumber
\end{eqnarray}
where the $C_{i}$ are constants we have used the exponential as a paradigmatic example of an entire function.\\
This gives us the following Lagrangian
\begin{eqnarray}
\label{vectorentire}
\mathcal{L}=\mathcal{L}_{M}+\frac{1}{6}C_{1}\hat{S}_{\mu}\hat{S}^{\mu}+\frac{1}{9}C_{2}\partial_{\left[\mu\right.}\hat{S}_{\left.\nu\right]}\partial^{\left[\mu\right.}\hat{S}^{\left.\nu\right]}+\frac{1}{3}C_{3}\hat{T}_{\mu}\hat{T}^{\mu}-\frac{2}{9}C_{4}\partial_{\left[\mu\right.}\hat{T}_{\left.\nu\right]}\partial^{\left[\mu\right.}\hat{T}^{\left.\nu\right]}-\frac{1}{9}C_{5}\partial_{\mu}\hat{T}^{\mu}\partial_{\nu}\hat{T}^{\nu},
\end{eqnarray}
where $\hat{S}^{\mu}=e^{-\frac{\Box}{2M_{S}^{2}}}S^{\mu}$ and $\hat{T}^{\mu}=e^{-\frac{\Box}{2M_{S}^{2}}}T^{\mu}$. From the standard theory of vector fields we know that the last term introduces ghostly degrees of freedom, therefore we need to impose that $C_{5}=0$. Moreover, the kinetic terms of both vectors need to be positive, hence we also have the conditions $C_{2}>0$ and $C_{4}<0$.

At this time we know that our theory is absent of ghosts, and we are ready to find some possible solutions, that we will show that can be singularity-free. We will study the solutions of the trace and axial vector separately in the following Subsections. This is indeed possible since parity breaking terms in the action are not considered, so there are no mixed trace-axial terms.

\subsubsection{Axial vector}
\label{axialsec}
First, we will consider the Cartan Equation for the axial vector \eqref{cartandecax}
\begin{eqnarray}
\label{cartananti2}
\frac{1}{6}C_{1}S_{\mu}+\frac{1}{18}C_{2}\left(\partial_{\mu}\partial_{\nu}S^{\nu}-\Box S_{\mu}\right)=e^{\frac{\Box}{M_{S}^{2}}}A_{\mu},
\end{eqnarray}
where $A_{\mu}=\frac{\delta\mathcal{L}_{fermion}}{\delta S^{\mu}}$ represents the internal spin of the fermion, that minimally couples to the axial vector \cite{Shapiro:2001rz}.\\

We realise that Eq.\eqref{cartananti2} for the axial vector, $S^{\mu}$, is very similar to the one in Ref. \cite{delaCruz-Dombriz:2018aal}, and can be solved analogously. Concretly, the Equations are the same if we impose $C_{1}=0$ and choose the gauge $\partial_{\mu}S^{\mu}=0$. Hence, it provides a non-singular solution for the axial torsion, see Ref. \cite{delaCruz-Dombriz:2018aal} for details of the derivation. More specifically, if we assume that the radial component of the axial vector is zero, we will then find that a spherically symmetric solution for a rotating ring singularity is indeed regularised, as found in \cite{delaCruz-Dombriz:2018aal}
\begin{eqnarray}
\label{solax}
&&S^{\mu}\left(\rho\right)=\frac{3}{2}A^{\mu}\int_{0}^{\infty}{\rm d}\xi{\rm J}_{0}\left(-R\xi\right){\rm J}_{0}\left(-\xi\rho\right){\rm Erfc}\left(\xi/{M_{s}}\right), \;\;\;\;\mu\neq r,
\\
&&S^{r}\left(\rho\right)={\rm constant},
\end{eqnarray}
where $A^{\mu}$ is constant, ${\rm J}_{0}$ is the Bessel function and ${\rm Erfc}(z)=1-{\rm Erf}(z)$ is the complementary error function.


\subsubsection{Trace vector}
\label{tracesec}

Let us now explore the Cartan Equation for the trace vector \eqref{cartandectra}
\begin{eqnarray}
\label{procatrace}
\frac{1}{3}C_{3}T_{\mu}-\frac{1}{9}C_{4}\left(\partial_{\mu}\partial_{\nu}T^{\nu}-\Box T_{\mu}\right)=0.
\end{eqnarray}
We observe that this is just the local Proca Equation for a vector field. Therefore, it will have the same plane wave solutions propagating three stable degrees of freedom.\\

Now, with all the components for the torsion tensor calculated, we will solve Einstein's equations to obtain the corresponding metric $h_{\mu\nu}$.

\subsection{Einstein Equations}

Let us recall that Einstein's equations for a fermionic source, where the tensor component of the torsion has been set to zero are given by \eqref{einsteindecomposed}:
\begin{eqnarray}
&&\Box a\left(\Box\right)h_{\mu\nu}+b\left(\Box\right)\partial_{\sigma}\partial_{\left(\nu\right.}h_{\left.\mu\right)}^{\,\,\,\sigma}+c\left(\Box\right)\left(\partial_{\mu}\partial_{\nu}h+\eta_{\mu\nu}\partial_{\rho}\partial_{\sigma}h^{\rho\sigma}\right)+\eta_{\mu\nu}\Box d\left(\Box\right)h+2\frac{f\left(\Box\right)}{\Box}\partial_{\mu}\partial_{\nu}\partial_{\rho}\partial_{\sigma}h^{\rho\sigma}=\tau_{\mu\nu},
\end{eqnarray}
where $\tau_{\mu\nu}=\eta_{\sigma\nu}F_{\mu\rho}F^{\sigma\rho}-\frac{1}{4}\eta_{\mu\nu}F_{\sigma\rho}F^{\sigma\rho}$, $F_{\mu\nu}$ being the electromagnetic tensor. It is clear that this equation is the same as in the pure metric case, since the torsion terms do not contribute.\\
Now, if we apply the constraints that we obtained from the energy-momentum conservation, and ghost-free conditions in the metric sector, see Eq.\eqref{constraints}, we are left with the following expression
\begin{eqnarray}
\label{einssimp}
{\rm {e}}^{-\Box/M_{s}^{2}}\left(\Box h_{\mu\nu}+\partial_{\mu}\partial_{\nu}h+\eta_{\mu\nu}\partial_{\rho}\partial_{\sigma}h^{\rho\sigma}-2\partial_{\sigma}\partial_{\left(\nu\right.}h_{\left.\mu\right)}^{\,\,\,\sigma}-\eta_{\mu\nu}\Box h\right)=\tau_{\mu\nu}.
\end{eqnarray}
It is interesting to note that this equation has already been studied in Ref.~\cite{Buoninfante:2018stt}, where a non-singular Reissner-Nordstr\"om solution were obtained for the same choice of the entire function in ghost free IDG, namely
\begin{equation}
{\rm d}s^{2}=-\left(1+2\Phi\left(r\right)\right){\rm d}t^{2}+\left(1-2\Psi\left(r\right)\right)\left({\rm d}r^{2}+r^{2}{\rm d}\Omega^{2}\right),
\end{equation}
where $\Phi\left(r\right)$ and $\Psi\left(r\right)$ take the following form \cite{Buoninfante:2018stt}
\begin{eqnarray}
\label{metricsol1}
&&\Phi\left(r\right)=-\frac{Gm}{r}\text{Erf}\left(\frac{M_{S}r}{2}\right)+\frac{GQ^{2}M_{S}}{2r}\text{F}\left(\frac{M_{S}r}{2}\right),
\\
&&\Psi\left(r\right)=-\frac{Gm}{r}\text{Erf}\left(\frac{M_{S}r}{2}\right)+\frac{GQ^{2}M_{S}}{4r}\text{F}\left(\frac{M_{S}r}{2}\right),
\label{metricsol2}
\end{eqnarray}
in which Erf$(x)$ is the error function and F$(x)$ the Dawson function.\
This solution is non-singular when $r\rightarrow 0$ and recasts a Reissner-Nordstr\"om  when $r\gg M_s^{-1}$.

\section{Final remarks}
\label{sec:VI}, 

We have provided the foundations of the theory of gravitation that can be constructed out infinite covariant derivatives and non-symmetric connection.The main advantages of this theory are the fact that one can introduce effective quantum effects, such as non-locality and internal spin of the particles, which ammeliorate the ring singularities present in the local theory, while preserving the stability of the spacetime. The disadvantage of course is that the calculations are quite tedious compared with GR ones. This issue can be solved if one sticks to the linear level, in which we have shown that solutions that are ghost and singularity free can be found, even with a non-vanishing torsion tensor. The method that we have used to obtain solutions is based on the decomposition of the torsion tensor into its Lorentz invariants, in particular the trace and axial vectors and the tensor part. We have assumed the latter to be zero due to symmetry arguments. Then, we have obtained the field equations of these two vectors and solve them for a fermionic source, see Eqs. \eqref{solax} and \eqref{procatrace}, finding the ghost and singularity-free conditions.

We have shown that in Einstein equations the torsion vectors decouple from the metric under the stability conditions, hence obtaining the same metric solutions as in the case of a torsion-free IDG, see Eqs. \eqref{metricsol1} and \eqref{metricsol2}, that are non-singular and free of ghosts.
Nevertheless, since the axial part of the torsion is different from zero, the phenomenology of the solution would be different to the one in the null torsion case \cite{Buoninfante:2018stt}, despite sharing the same metric solution. This is because totally antisymmetric part of the torsion, i.e. the axial vector, couples with the internal spin of the fermionic source, which produces a non-geodesical behavior in the fermions, that it is not observed when torsion is set zero (see \cite{Cembranos:2018ipn} for details). 

For future work in this theory, it will be interesting to calculate the next to leading order of the field equations, so that one can find torsion effects in the effective energy-momentum tensor of Einstein's equations, allowing us to make the UV extension of Poincar\'e gauge solutions. On the other hand, the influence of torsion and non-locality in quantum experiments remains of interest, so experimental constraints on the viability of the theory could be provided, see ref.~\cite{Bose:2017nin}.

\acknowledgements
 AM's research is funded by the Netherlands Organization for Scientific Research (NWO) grant number 680-91-119. AdlCD and FJMT acknowledge financial support from UCT Launching Grants Programme and NRF Grants No. 99077 2016-2018, Ref. No. CSUR150628121624, 110966 Ref. No. BS170509230233, and the NRF Incentive Funding for Rated Researchers (IPRR), Ref. No. IFR170131220846. FJMT acknowledges financial support from the Erasmus+ KA107 Alliance4Universities programme and from the Van Swinderen Institute at the University of Groningen. AdlCD acknowledges financial support from Project No. FPA2014-53375-C2-1-P from the Spanish Ministry of Economy and Science, Project No. FIS2016-78859-P from the European Regional Development Fund and Spanish Research Agency (AEI), and Project No. CA16104 from COST Action EU Framework Programme Horizon 2020. 
AdlCD and FJMT thank the hospitality of the Institute of Theoretical Astrophysics - University of Oslo (Norway) during the latter steps of the manuscript.

\appendix

\section{Components of the action}
\label{ap1}

In this Appendix we give the different terms that appear in the linearised action \eqref{lagrangian}.

\begin{eqnarray}
\tilde{R}\tilde{F}_{1}\left(\Box\right)\tilde{R}&=&\tilde{F}_{1}\left(\Box\right)\left[h\Box^{2}h+h^{\rho\sigma}\partial_{\rho}\partial_{\sigma}\partial_{\mu}\partial_{\nu}h^{\mu\nu}-2h\Box\partial_{\mu}\partial_{\nu}h^{\mu\nu}-4h^{\mu\nu}\partial_{\mu}\partial_{\nu}\partial_{\rho}K_{\,\,\,\,\,\sigma}^{\rho\sigma}\right.
\nonumber
\\
&+&\left. 4h\Box\partial_{\rho}K_{\,\,\,\,\,\sigma}^{\rho\sigma}-4K_{\,\,\,\,\,\sigma}^{\rho\sigma}\partial_{\rho}\partial_{\mu}K_{\,\,\,\,\,\nu}^{\mu\nu}\right],
\end{eqnarray}

\begin{eqnarray}
\tilde{R}\tilde{F}_{2}\left(\Box\right)\partial_{\mu}\partial_{\nu}\tilde{R}^{\mu\nu}&=&\tilde{F}_{2}\left(\Box\right)\left[\frac{1}{2}h^{\rho\sigma}\Box\partial_{\rho}\partial_{\sigma}\partial_{\mu}\partial_{\nu}h^{\mu\nu}-h\Box^{2}\partial_{\mu}\partial_{\nu}h^{\mu\nu}+\frac{1}{2}h\Box^{3}h\right.
\nonumber
\\
&-&\left.h^{\mu\nu}\Box\partial_{\mu}\partial_{\nu}\partial_{\rho}K_{\,\,\,\,\,\sigma}^{\rho\sigma}-2K_{\,\,\,\,\,\sigma}^{\rho\sigma}\Box\partial_{\rho}\partial_{\nu}K_{\,\,\,\,\,\lambda}^{\nu\lambda}\right],
\end{eqnarray}

\begin{eqnarray}
\tilde{R}_{\mu\nu}\tilde{F}_{3}\left(\Box\right)\tilde{R}^{\left(\mu\nu\right)}&=&\tilde{F}_{3}\left(\Box\right)\left[\frac{1}{4}h\Box^{2}h+\frac{1}{4}h_{\mu\nu}\Box^{2}h^{\mu\nu}-\frac{1}{2}h_{\mu}^{\sigma}\partial_{\sigma}\partial_{\nu}h^{\mu\nu}-\frac{1}{2}h\Box\partial_{\mu}\partial_{\nu}h^{\mu\nu}\right.
\nonumber
\\
&+&\frac{1}{2}h^{\mu\nu}\partial_{\sigma}\partial_{\mu}\partial_{\nu}\partial_{\rho}h^{\rho\sigma}-\frac{1}{2}h_{\mu}^{\sigma}\partial_{\sigma}\partial_{\nu}\partial_{\rho}K^{\rho\nu\mu}-\frac{1}{2}h^{\nu\sigma}\partial_{\sigma}\partial_{\nu}\partial_{\mu}K_{\,\,\,\,\,\rho}^{\mu\rho}-\frac{1}{2}h_{\mu}^{\sigma}\partial_{\sigma}\Box K_{\,\,\,\,\,\rho}^{\mu\rho}
\nonumber
\\
&+&\left.\frac{1}{2}h_{\mu\nu}\Box\partial_{\rho}K^{\rho\mu\nu}-K_{\,\,\mu\nu}^{\rho}\partial_{\rho}\partial_{\sigma}K^{\sigma\left(\mu\nu\right)}-K_{\,\,\mu\nu}^{\rho}\partial_{\rho}\partial^{\mu}K_{\,\,\,\,\,\lambda}^{\nu\lambda}-\frac{1}{2}K_{\,\,\,\,\,\lambda}^{\nu\lambda}\Box K_{\nu\,\,\,\rho}^{\,\,\rho}-\frac{1}{2}K_{\,\,\,\,\,\rho}^{\mu\rho}\partial_{\mu}\partial_{\nu}K_{\,\,\,\,\,\lambda}^{\nu\lambda}\right],
\end{eqnarray}

\begin{equation}
\tilde{R}_{\mu\nu}\tilde{F}_{4}\left(\Box\right)\tilde{R}^{\left[\mu\nu\right]}=\tilde{F}_{4}\left(\Box\right)\left[-K_{\,\,\mu\nu}^{\rho}\partial_{\rho}\partial_{\sigma}K^{\sigma\left[\mu\nu\right]}-K_{\,\,\mu\nu}^{\rho}\partial_{\rho}\partial^{\mu}K_{\,\,\,\,\,\lambda}^{\nu\lambda}-\frac{1}{2}K_{\,\,\,\,\,\lambda}^{\nu\lambda}\Box K_{\nu\,\,\,\rho}^{\,\,\rho}+\frac{1}{2}K_{\,\,\,\,\,\rho}^{\mu\rho}\partial_{\mu}\partial_{\nu}K_{\,\,\,\,\,\lambda}^{\nu\lambda}\right],
\end{equation}

\begin{eqnarray}
\tilde{R}_{\left(\mu\right.}^{\,\,\,\left.\nu\right)}\tilde{F}_{5}\left(\Box\right)\partial_{\nu}\partial_{\lambda}\tilde{R}^{\mu\lambda}&=&\tilde{F}_{5}\left(\Box\right)\left[\frac{1}{4}h\Box^{3}h-\frac{1}{2}h\Box^{2}\partial_{\mu}\partial_{\nu}h^{\mu\nu}+\frac{1}{4}h^{\lambda\sigma}\Box\partial_{\sigma}\partial_{\lambda}\partial_{\mu}\partial_{\nu}h^{\mu\nu}\right.
\nonumber
\\
&-&\left.h^{\nu\sigma}\Box\partial_{\sigma}\partial_{\nu}\partial_{\mu}K_{\,\,\,\,\,\rho}^{\mu\rho}+h\Box^{2}\partial_{\nu}K_{\,\,\,\,\,\lambda}^{\nu\lambda}-K_{\,\,\,\,\,\rho}^{\mu\rho}\Box\partial_{\mu}\partial_{\nu}K_{\,\,\,\,\,\lambda}^{\nu\lambda}\right],
\end{eqnarray}

\begin{equation}
\tilde{R}_{\left[\mu\right.}^{\,\,\,\left.\nu\right]}\tilde{F}_{6}\left(\Box\right)\partial_{\nu}\partial_{\lambda}\tilde{R}^{\mu\lambda}=0,
\end{equation}

\begin{eqnarray}
\tilde{R}_{\mu}^{\,\,\,\nu}\tilde{F}_{7}\left(\Box\right)\partial_{\nu}\partial_{\lambda}\tilde{R}^{\left(\mu\lambda\right)}&=&\tilde{F}_{7}\left(\Box\right)\left[\frac{1}{4}h\Box^{3}h-\frac{1}{2}h\Box^{2}\partial_{\mu}\partial_{\nu}h^{\mu\nu}+\frac{1}{4}h^{\lambda\sigma}\Box\partial_{\sigma}\partial_{\lambda}\partial_{\mu}\partial_{\nu}h^{\mu\nu}-h^{\nu\sigma}\Box\partial_{\sigma}\partial_{\nu}\partial_{\mu}K_{\,\,\,\,\,\rho}^{\mu\rho}\right.
\nonumber
\\
&+&\left.h\Box^{2}\partial_{\nu}K_{\,\,\,\,\,\lambda}^{\nu\lambda}-K_{\,\,\,\,\,\rho}^{\mu\rho}\Box\partial_{\mu}\partial_{\nu}K_{\,\,\,\,\,\lambda}^{\nu\lambda}\right],
\end{eqnarray}

\begin{equation}
\tilde{R}_{\mu}^{\,\,\,\nu}\tilde{F}_{8}\left(\Box\right)\partial_{\nu}\partial_{\lambda}\tilde{R}^{\left[\mu\lambda\right]}=0,
\end{equation}

\begin{eqnarray}
\tilde{R}^{\lambda\sigma}\tilde{F}_{9}\left(\Box\right)\partial_{\mu}\partial_{\sigma}\partial_{\nu}\partial_{\lambda}\tilde{R}^{\mu\nu}&=&\tilde{F}_{9}\left(\Box\right)\left[\frac{1}{4}h\Box^{4}h-\frac{1}{2}h\Box^{3}\partial_{\mu}\partial_{\nu}h^{\mu\nu}+\frac{1}{4}h^{\lambda\sigma}\Box^{2}\partial_{\sigma}\partial_{\lambda}\partial_{\mu}\partial_{\nu}h^{\mu\nu}-h^{\nu\sigma}\Box^{2}\partial_{\sigma}\partial_{\nu}\partial_{\mu}K_{\,\,\,\,\,\rho}^{\mu\rho}\right.
\nonumber
\\
&+&\left.h\Box^{3}\partial_{\nu}K_{\,\,\,\,\,\lambda}^{\nu\lambda}-K_{\,\,\,\,\,\rho}^{\mu\rho}\Box^{2}\partial_{\mu}\partial_{\nu}K_{\,\,\,\,\,\lambda}^{\nu\lambda}\right],
\end{eqnarray}

\begin{eqnarray}
\tilde{R}_{\left(\mu\lambda\right)}\tilde{F}_{10}\left(\Box\right)\partial_{\nu}\partial_{\sigma}\tilde{R}^{\mu\nu\lambda\sigma}&=&\tilde{F}_{10}\left(\Box\right)\left[\frac{1}{4}h_{\mu\lambda}\Box^{3}h^{\mu\lambda}-\frac{1}{2}h_{\mu}^{\,\,\alpha}\Box^{2}\partial_{\alpha}\partial_{\sigma}h^{\sigma\mu}+\frac{1}{4}h^{\lambda\sigma}\Box\partial_{\sigma}\partial_{\lambda}\partial_{\mu}\partial_{\nu}h^{\mu\nu}\right.
\nonumber
\\
&-&h_{\mu\sigma}\Box^{2}\partial_{\lambda}K^{\mu\sigma\lambda}+h_{\mu}^{\,\,\alpha}\Box\partial_{\alpha}\partial_{\lambda}\partial_{\sigma}K^{\mu\sigma\lambda}+K_{\alpha\left(\mu\lambda\right)}\Box\partial^{\alpha}\partial_{\sigma}K^{\lambda\mu\sigma}
\nonumber
\\
&-&\left.\frac{1}{2}K_{\alpha\mu\lambda}\partial^{\alpha}\partial^{\mu}\partial_{\sigma}\partial_{\nu}K^{\lambda\nu\sigma}\right],
\end{eqnarray}

\begin{equation}
\tilde{R}_{\left[\mu\lambda\right]}\tilde{F}_{11}\left(\Box\right)\partial_{\nu}\partial_{\sigma}\tilde{R}^{\mu\nu\lambda\sigma}=\tilde{F}_{11}\left(\Box\right)\left[K_{\alpha\left[\mu\lambda\right]}\Box\partial^{\alpha}\partial_{\sigma}K^{\lambda\mu\sigma}-\frac{1}{2}K_{\alpha\mu\lambda}\partial^{\alpha}\partial^{\mu}\partial_{\sigma}\partial_{\nu}K^{\lambda\nu\sigma}\right],
\end{equation}

\begin{eqnarray}
\tilde{R}_{\mu\lambda}\tilde{F}_{12}\left(\Box\right)\partial_{\nu}\partial_{\sigma}\left(\tilde{R}^{\mu\nu\lambda\sigma}+\tilde{R}^{\lambda\sigma\mu\nu}\right)&=&\tilde{F}_{12}\left(\Box\right)\left[\frac{1}{2}h_{\mu\lambda}\Box^{3}h^{\mu\lambda}-h_{\mu}^{\,\,\alpha}\Box^{2}\partial_{\alpha}\partial_{\sigma}h^{\sigma\mu}+\frac{1}{2}h^{\lambda\sigma}\Box\partial_{\sigma}\partial_{\lambda}\partial_{\mu}\partial_{\nu}h^{\mu\nu}\right.
\nonumber
\\
&+&2h_{\mu}^{\,\,\alpha}\Box\partial_{\alpha}\partial_{\lambda}\partial_{\sigma}K^{\mu\sigma\lambda}-2h_{\mu\sigma}\Box^{2}\partial_{\lambda}K^{\mu\sigma\lambda}+2K_{\alpha\left(\mu\lambda\right)}\Box\partial^{\alpha}\partial_{\sigma}K^{\lambda\mu\sigma}
\nonumber
\\
&-&\left.K_{\alpha\mu\lambda}\partial^{\alpha}\partial^{\mu}\partial_{\sigma}\partial_{\nu}K^{\lambda\nu\sigma}\right],
\end{eqnarray}

\begin{equation}
\tilde{R}_{\mu\lambda}\tilde{F}_{13}\left(\Box\right)\partial_{\nu}\partial_{\sigma}\left(\tilde{R}^{\mu\nu\lambda\sigma}-\tilde{R}^{\lambda\sigma\mu\nu}\right)=\tilde{F}_{13}\left(\Box\right)\left[2K_{\alpha\left[\mu\lambda\right]}\Box\partial^{\alpha}\partial_{\sigma}K^{\lambda\mu\sigma}-K_{\alpha\mu\lambda}\partial^{\alpha}\partial^{\mu}\partial_{\sigma}\partial_{\nu}K^{\lambda\nu\sigma}\right],
\end{equation}

\begin{eqnarray}
\tilde{R}_{\mu\nu\lambda\sigma}\tilde{F}_{14}\left(\Box\right)\left(\tilde{R}^{\mu\nu\lambda\sigma}+\tilde{R}^{\lambda\sigma\mu\nu}\right)&=&\tilde{F}_{14}\left(\Box\right)\left[2h_{\mu\lambda}\Box^{2}h^{\mu\lambda}+2h^{\lambda\sigma}\partial_{\sigma}\partial_{\lambda}\partial_{\mu}\partial_{\nu}h^{\mu\nu}-4h_{\mu}^{\,\,\alpha}\Box\partial_{\alpha}\partial_{\sigma}h^{\sigma\mu}\right.
\nonumber
\\
&+&8h_{\sigma\mu}\Box\partial_{\nu}K^{\nu\mu\sigma}+8h_{\sigma\mu}\partial_{\nu}\partial_{\lambda}\partial^{\mu}K^{\sigma\lambda\nu}-2K^{\mu\sigma\lambda}\Box K_{\mu\sigma\lambda}
\nonumber
\\
&-&\left.4K^{\nu\sigma\lambda}\partial_{\nu}\partial^{\mu}K_{\mu\lambda\sigma}+2K^{\lambda\nu\mu}\partial_{\nu}\partial^{\sigma}K_{\lambda\sigma\mu}\right],
\end{eqnarray}

\begin{eqnarray}
\tilde{R}_{\mu\nu\lambda\sigma}\tilde{F}_{15}\left(\Box\right)\left(\tilde{R}^{\mu\nu\lambda\sigma}-\tilde{R}^{\lambda\sigma\mu\nu}\right)=\tilde{F}_{15}\left(\Box\right)\left[-2K^{\mu\sigma\lambda}\Box K_{\mu\sigma\lambda}+4K^{\nu\sigma\lambda}\partial_{\nu}\partial^{\mu}K_{\mu\lambda\sigma}+2K^{\lambda\nu\mu}\partial_{\nu}\partial^{\sigma}K_{\lambda\sigma\mu}\right],
\end{eqnarray}

\begin{eqnarray}
\left(\tilde{R}_{\rho\mu\nu\lambda}+\tilde{R}_{\nu\lambda\rho\mu}\right)\tilde{F}_{16}\left(\Box\right)\partial^{\rho}\partial_{\sigma}\tilde{R}^{\mu\nu\lambda\sigma}&=&\tilde{F}_{16}\left(\Box\right)\left[\frac{1}{2}h_{\mu\lambda}\Box^{3}h^{\mu\lambda}-h_{\mu}^{\,\,\alpha}\Box^{2}\partial_{\alpha}\partial_{\sigma}h^{\sigma\mu}+\frac{1}{2}h^{\lambda\sigma}\Box\partial_{\sigma}\partial_{\lambda}\partial_{\mu}\partial_{\nu}h^{\mu\nu}\right.
\nonumber
\\
&+&2h_{\sigma\mu}\Box^{2}\partial_{\nu}K^{\nu\mu\sigma}+2h_{\sigma\mu}\Box\partial_{\nu}\partial_{\lambda}\partial^{\mu}K^{\sigma\lambda\nu}+2K_{\alpha\left(\mu\lambda\right)}\Box\partial^{\alpha}\partial_{\sigma}K^{\lambda\mu\sigma}
\nonumber
\\
&-&\left.K_{\alpha\mu\lambda}\partial^{\alpha}\partial^{\mu}\partial_{\sigma}\partial_{\nu}K^{\lambda\nu\sigma}\right],
\end{eqnarray}

\begin{equation}
\left(\tilde{R}_{\rho\mu\nu\lambda}-\tilde{R}_{\nu\lambda\rho\mu}\right)\tilde{F}_{17}\left(\Box\right)\partial^{\rho}\partial_{\sigma}\tilde{R}^{\mu\nu\lambda\sigma}=\tilde{F}_{17}\left(\Box\right)\left[-2K^{\mu\sigma\lambda}\Box\partial^{\rho}\partial_{\sigma}K_{\lambda\mu\rho}-2K^{\nu\sigma\lambda}\partial^{\mu}\partial^{\rho}\partial_{\sigma}\partial_{\lambda}K_{\nu\mu\rho}\right],
\end{equation}

\begin{eqnarray}
\tilde{R}_{\rho\mu\nu\lambda}\tilde{F}_{18}\left(\Box\right)\partial^{\rho}\partial_{\sigma}\left(\tilde{R}^{\mu\nu\lambda\sigma}+\tilde{R}^{\lambda\sigma\mu\nu}\right)&=&\tilde{F}_{18}\left(\Box\right)\left[\frac{1}{2}h_{\mu\lambda}\Box^{3}h^{\mu\lambda}-h_{\mu}^{\,\,\alpha}\Box^{2}\partial_{\alpha}\partial_{\sigma}h^{\sigma\mu}+\frac{1}{2}h^{\lambda\sigma}\Box\partial_{\sigma}\partial_{\lambda}\partial_{\mu}\partial_{\nu}h^{\mu\nu}\right.
\nonumber
\\
&+&2h_{\sigma\mu}\Box^{2}\partial_{\nu}K^{\nu\mu\sigma}+2h_{\sigma\mu}\Box\partial_{\nu}\partial_{\lambda}\partial^{\mu}K^{\sigma\lambda\nu}+2K_{\alpha\mu\lambda}\Box\partial^{\alpha}\partial_{\sigma}K^{\lambda\mu\sigma}
\nonumber
\\
&-&\left.2K_{\left[\nu\mu\right]\lambda}\Box\partial_{\sigma}\partial^{\lambda}K^{\mu\sigma\nu}+K^{\mu\sigma\lambda}\Box^{2}K_{\sigma\lambda\mu}\right],
\end{eqnarray}

\begin{eqnarray}
\tilde{R}_{\rho\mu\nu\lambda}\tilde{F}_{19}\left(\Box\right)\partial^{\rho}\partial_{\sigma}\left(\tilde{R}^{\mu\nu\lambda\sigma}-\tilde{R}^{\lambda\sigma\mu\nu}\right)&=&\tilde{F}_{19}\left(\Box\right)\left[-2K_{\alpha\mu\lambda}\partial^{\alpha}\partial^{\mu}\partial_{\sigma}\partial_{\nu}K^{\lambda\nu\sigma}+2K_{\left[\nu\mu\right]\lambda}\Box\partial_{\sigma}\partial^{\lambda}K^{\mu\sigma\nu}\right.
\nonumber
\\
&+&\left.K_{\alpha\mu\lambda}\Box\partial^{\alpha}\partial_{\sigma}K^{\lambda\mu\sigma}-K^{\mu\sigma\lambda}\Box^{2}K_{\sigma\lambda\mu}\right],
\end{eqnarray}

\begin{eqnarray}
\left(\tilde{R}_{\mu\nu\rho\sigma}+\tilde{R}_{\rho\sigma\mu\nu}\right)\tilde{F}_{20}\left(\Box\right)\partial^{\nu}\partial^{\sigma}\partial_{\alpha}\partial_{\beta}\tilde{R}^{\mu\alpha\rho\beta}&=&\tilde{F}_{20}\left(\Box\right)\left[\frac{1}{2}h_{\mu\lambda}\Box^{4}h^{\mu\lambda}-h_{\mu}^{\,\,\alpha}\Box^{3}\partial_{\alpha}\partial_{\sigma}h^{\sigma\mu}+\frac{1}{2}h^{\lambda\sigma}\Box^{2}\partial_{\sigma}\partial_{\lambda}\partial_{\mu}\partial_{\nu}h^{\mu\nu}\right.
\nonumber
\\
&+&2h_{\mu}^{\,\,\alpha}\Box^{2}\partial_{\alpha}\partial_{\lambda}\partial_{\sigma}K^{\mu\sigma\lambda}-2h_{\mu\lambda}\Box^{3}\partial_{\sigma}K^{\mu\lambda\sigma}+2K_{\alpha\left(\mu\lambda\right)}\Box\partial^{\alpha}\partial_{\sigma}K^{\lambda\mu\sigma}
\nonumber
\\
&-&\left.K_{\alpha\mu\lambda}\partial^{\alpha}\partial^{\mu}\partial_{\sigma}\partial_{\nu}K^{\lambda\nu\sigma}\right],
\end{eqnarray}

\begin{equation}
\left(\tilde{R}_{\mu\nu\rho\sigma}-\tilde{R}_{\rho\sigma\mu\nu}\right)\tilde{F}_{21}\left(\Box\right)\partial^{\nu}\partial^{\sigma}\partial_{\alpha}\partial_{\beta}\tilde{R}^{\mu\alpha\rho\beta}=\tilde{F}_{21}\left(\Box\right)\left[2K_{\alpha\left[\mu\lambda\right]}\Box\partial^{\alpha}\partial_{\sigma}K^{\mu\lambda\sigma}-K_{\alpha\mu\lambda}\partial^{\alpha}\partial^{\mu}\partial_{\sigma}\partial_{\nu}K^{\lambda\nu\sigma}\right],
\end{equation}

\begin{eqnarray}
\tilde{R}_{\mu\nu\rho\sigma}\tilde{F}_{22}\left(\Box\right)\partial^{\nu}\partial^{\sigma}\partial_{\alpha}\partial_{\beta}\left(\tilde{R}^{\mu\alpha\rho\beta}+\tilde{R}^{\rho\beta\mu\alpha}\right)&=&\tilde{F}_{22}\left(\Box\right)\left[\frac{1}{2}h_{\mu\lambda}\Box^{4}h^{\mu\lambda}-h_{\mu}^{\,\,\alpha}\Box^{3}\partial_{\alpha}\partial_{\sigma}h^{\sigma\mu}+\frac{1}{2}h^{\lambda\sigma}\Box^{2}\partial_{\sigma}\partial_{\lambda}\partial_{\mu}\partial_{\nu}h^{\mu\nu}\right.
\nonumber
\\
&+&2h_{\mu}^{\,\,\alpha}\Box^{2}\partial_{\alpha}\partial_{\lambda}\partial_{\sigma}K^{\mu\sigma\lambda}-2h_{\mu\lambda}\Box^{3}\partial_{\sigma}K^{\mu\lambda\sigma}+2K_{\alpha\left(\mu\lambda\right)}\Box\partial^{\alpha}\partial_{\sigma}K^{\lambda\mu\sigma}
\nonumber
\\
&-&\left.K_{\alpha\mu\lambda}\partial^{\alpha}\partial^{\mu}\partial_{\sigma}\partial_{\nu}K^{\lambda\nu\sigma}\right],
\end{eqnarray}

\begin{eqnarray}
\tilde{R}_{\mu\nu\rho\sigma}\tilde{F}_{23}\left(\Box\right)\partial^{\nu}\partial^{\sigma}\partial_{\alpha}\partial_{\beta}\left(\tilde{R}^{\mu\alpha\rho\beta}-\tilde{R}^{\rho\beta\mu\alpha}\right)&=&\tilde{F}_{23}\left(\Box\right)\left[-2K_{\alpha\left[\mu\lambda\right]}\Box\partial^{\alpha}\partial_{\sigma}K^{\mu\lambda\sigma}-K_{\alpha\mu\lambda}\partial^{\alpha}\partial^{\mu}\partial_{\sigma}\partial_{\nu}K^{\lambda\nu\sigma}\right],
\end{eqnarray}

\begin{eqnarray}
\tilde{R}_{\mu\nu\rho\sigma}\tilde{F}_{34}\left(\Box\right)\partial^{\mu}K^{\nu\rho\sigma}=\tilde{F}_{34}\left(\Box\right)\left[-\frac{1}{2}h_{\sigma\mu}\partial_{\nu}\partial_{\lambda}\partial^{\mu}K^{\sigma\lambda\nu}-\frac{1}{2}h_{\sigma\mu}\Box\partial_{\nu}K^{\nu\mu\sigma}+K_{\nu\mu\lambda}\partial_{\sigma}\partial^{\lambda}K^{\mu\sigma\nu}-K^{\mu\sigma\lambda}\Box K_{\sigma\mu\lambda}\right],
\end{eqnarray}

\begin{eqnarray}
\tilde{R}_{\mu\nu\rho\sigma}\tilde{F}_{35}\left(\Box\right)\partial^{\rho}K^{\mu\nu\sigma}=\tilde{F}_{35}\left(\Box\right)\left[-\frac{1}{2}h_{\sigma\mu}\partial_{\nu}\partial_{\lambda}\partial^{\mu}K^{\sigma\lambda\nu}-\frac{1}{2}h_{\sigma\mu}\Box\partial_{\nu}K^{\nu\mu\sigma}-K_{\nu\mu\lambda}\partial_{\sigma}\partial^{\lambda}K^{\mu\sigma\nu}+K_{\lambda\mu\nu}\partial_{\sigma}\partial^{\lambda}K^{\sigma\mu\nu}\right],
\end{eqnarray}

\begin{eqnarray}
\tilde{R}_{\left(\rho\sigma\right)}\tilde{F}_{36}\left(\Box\right)\partial_{\nu}K^{\mu\nu\sigma}=\tilde{F}_{36}\left(\Box\right)\left[-\frac{1}{2}h_{\sigma\mu}\partial_{\nu}\partial_{\lambda}\partial^{\mu}K^{\sigma\lambda\nu}-\frac{1}{2}h_{\sigma\mu}\Box\partial_{\nu}K^{\nu\mu\sigma}+K_{\lambda\left(\mu\nu\right)}\partial_{\sigma}\partial^{\lambda}K^{\sigma\mu\nu}-\frac{1}{2}K_{\,\,\,\lambda\sigma}^{\lambda}\partial_{\rho}\partial_{\nu}K^{\nu\rho\sigma}\right],
\end{eqnarray}

\begin{eqnarray}
\tilde{R}_{\left[\rho\sigma\right]}\tilde{F}_{37}\left(\Box\right)\partial_{\nu}K^{\nu\rho\sigma}=\tilde{F}_{37}\left(\Box\right)\left[K_{\lambda\left[\mu\nu\right]}\partial_{\sigma}\partial^{\lambda}K^{\sigma\mu\nu}-\frac{1}{2}K_{\,\,\,\lambda\sigma}^{\lambda}\partial_{\rho}\partial_{\nu}K^{\nu\rho\sigma}\right],
\end{eqnarray}

\begin{eqnarray}
\tilde{R}_{\rho\sigma}\tilde{F}_{38}\left(\Box\right)\partial_{\nu}K^{\rho\nu\sigma}=\tilde{F}_{38}\left(\Box\right)\left[-K_{\nu\mu\lambda}\partial_{\sigma}\partial^{\lambda}K^{\mu\sigma\nu}-K_{\,\,\,\lambda\sigma}^{\lambda}\partial_{\rho}\partial_{\nu}K^{\nu\rho\sigma}\right],
\end{eqnarray}

\begin{eqnarray}
\tilde{R}_{\left(\rho\sigma\right)}\tilde{F}_{39}\left(\Box\right)\partial^{\sigma}K_{\,\,\,\,\,\,\mu}^{\rho\mu}&=&\tilde{F}_{39}\left(\Box\right)\left[\frac{1}{2}h_{\sigma\lambda}\partial^{\sigma}\partial^{\lambda}\partial_{\rho}K_{\,\,\,\,\,\,\mu}^{\rho\mu}-\frac{1}{2}h\Box\partial_{\rho}K_{\,\,\,\,\,\,\mu}^{\rho\mu}+\frac{1}{2}K_{\nu\mu\rho}\partial^{\nu}\partial^{\mu}K_{\,\,\,\,\,\,\mu}^{\rho\mu}-\frac{1}{2}K_{\,\,\,\lambda\sigma}^{\lambda}\partial^{\sigma}\partial_{\rho}K_{\,\,\,\,\,\,\mu}^{\rho\mu}\right.
\nonumber
\\
&-&\left.\frac{1}{2}K_{\,\,\,\lambda\rho}^{\lambda}\Box K_{\,\,\,\,\,\,\mu}^{\rho\mu}\right],
\end{eqnarray}

\begin{eqnarray}
\tilde{R}_{\left[\rho\sigma\right]}\tilde{F}_{40}\left(\Box\right)\partial^{\sigma}K_{\,\,\,\,\,\,\mu}^{\rho\mu}=\tilde{F}_{40}\left(\Box\right)\left[-\frac{1}{2}K_{\nu\mu\rho}\partial^{\nu}\partial^{\mu}K_{\,\,\,\,\,\,\mu}^{\rho\mu}-\frac{1}{2}K_{\,\,\,\lambda\sigma}^{\lambda}\partial^{\sigma}\partial_{\rho}K_{\,\,\,\,\,\,\mu}^{\rho\mu}+\frac{1}{2}K_{\,\,\,\lambda\rho}^{\lambda}\Box K_{\,\,\,\,\,\,\mu}^{\rho\mu}\right],
\end{eqnarray}

\begin{eqnarray}
\tilde{R}\tilde{F}_{41}\left(\Box\right)\partial_{\rho}K_{\,\,\,\,\,\,\mu}^{\rho\mu}=\tilde{F}_{41}\left(\Box\right)\left[h_{\sigma\lambda}\partial^{\sigma}\partial^{\lambda}\partial_{\rho}K_{\,\,\,\,\,\,\mu}^{\rho\mu}-h\Box\partial_{\rho}K_{\,\,\,\,\,\,\mu}^{\rho\mu}-2K_{\,\,\,\lambda\sigma}^{\lambda}\partial^{\sigma}\partial_{\rho}K_{\,\,\,\,\,\,\mu}^{\rho\mu}\right],
\end{eqnarray}

\begin{eqnarray}
\tilde{R}_{\mu\alpha\rho\sigma}\tilde{F}_{42}\left(\Box\right)\partial^{\mu}\partial^{\rho}\partial_{\nu}K^{\nu\left(\alpha\sigma\right)}&=&\tilde{F}_{42}\left(\Box\right)\left[-\frac{1}{2}h_{\sigma\mu}\Box\partial_{\nu}\partial_{\lambda}\partial^{\mu}K^{\sigma\lambda\nu}-\frac{1}{2}h_{\sigma\mu}\Box^{2}\partial_{\nu}K^{\nu\mu\sigma}-\frac{1}{2}K_{\nu\mu\lambda}\partial_{\alpha}\partial_{\rho}\partial^{\nu}\partial^{\mu}K^{\alpha\rho\lambda}\right.
\nonumber
\\
&+&\left.K_{\lambda\left(\mu\nu\right)}\Box\partial_{\sigma}\partial^{\lambda}K^{\sigma\mu\nu}\right],
\end{eqnarray}

\begin{eqnarray}
\tilde{R}_{\mu\alpha\rho\sigma}\tilde{F}_{43}\left(\Box\right)\partial^{\mu}\partial^{\rho}\partial_{\nu}K^{\nu\left[\alpha\sigma\right]}=\tilde{F}_{43}\left(\Box\right)\left[-\frac{1}{2}K_{\nu\mu\lambda}\partial_{\alpha}\partial_{\rho}\partial^{\nu}\partial^{\mu}K^{\alpha\rho\lambda}+K_{\lambda\left[\mu\nu\right]}\Box\partial_{\sigma}\partial^{\lambda}K^{\sigma\mu\nu}\right],
\end{eqnarray}

\begin{eqnarray}
\tilde{R}_{\mu\alpha\rho\sigma}\tilde{F}_{44}\left(\Box\right)\partial^{\mu}\partial^{\rho}\partial_{\nu}K^{\alpha\nu\sigma}=\tilde{F}_{44}\left(\Box\right)\left[-K_{\nu\mu\lambda}\partial_{\alpha}\partial_{\rho}\partial^{\nu}\partial^{\mu}K^{\alpha\rho\lambda}+K_{\lambda\mu\nu}\Box\partial_{\sigma}\partial^{\lambda}K^{\mu\sigma\nu}\right],
\end{eqnarray}

\begin{eqnarray}
\tilde{R}_{\left(\rho\sigma\right)}\tilde{F}_{45}\left(\Box\right)\partial^{\rho}\partial_{\nu}\partial_{\alpha}K^{\sigma\nu\alpha}=\tilde{F}_{45}\left(\Box\right)\left[-\frac{1}{2}K_{\nu\mu\lambda}\partial_{\alpha}\partial_{\rho}\partial^{\nu}\partial^{\mu}K^{\alpha\rho\lambda}-\frac{1}{2}K_{\,\,\,\lambda\sigma}^{\lambda}\Box\partial_{\mu}\partial_{\alpha}K^{\sigma\mu\alpha}\right],
\end{eqnarray}

\begin{eqnarray}
\tilde{R}_{\left(\rho\sigma\right)}\tilde{F}_{46}\left(\Box\right)\partial^{\rho}\partial_{\nu}\partial_{\alpha}K^{\sigma\nu\alpha}=\tilde{F}_{46}\left(\Box\right)\left[-\frac{1}{2}K_{\nu\mu\lambda}\partial_{\alpha}\partial_{\rho}\partial^{\nu}\partial^{\mu}K^{\alpha\rho\lambda}-\frac{1}{2}K_{\,\,\,\lambda\sigma}^{\lambda}\Box\partial_{\mu}\partial_{\alpha}K^{\sigma\mu\alpha}\right],
\end{eqnarray}

\begin{eqnarray}
\tilde{R}_{\mu\nu\lambda\sigma}\tilde{F}_{47}\left(\Box\right)\widetilde{R}^{\mu\lambda\nu\sigma}&=&\tilde{F}_{47}\left(\Box\right)\left[h_{\mu\lambda}\Box^{2}h^{\mu\lambda}+h^{\lambda\sigma}\partial_{\sigma}\partial_{\lambda}\partial_{\mu}\partial_{\nu}h^{\mu\nu}-2h_{\mu}^{\,\,\alpha}\Box\partial_{\alpha}\partial_{\sigma}h^{\sigma\mu}\right.
\nonumber
\\
&+&4h_{\sigma\mu}\Box\partial_{\nu}K^{\nu\mu\sigma}+4h_{\sigma\mu}\partial_{\nu}\partial_{\lambda}\partial^{\mu}K^{\sigma\lambda\nu}-K^{\mu\sigma\lambda}\Box K_{\mu\lambda\sigma}
\nonumber
\\
&-&\left.K^{\nu\lambda\sigma}\partial_{\nu}\partial^{\mu}K_{\mu\lambda\sigma}+2K^{\lambda\nu\mu}\partial_{\nu}\partial^{\sigma}K_{\lambda\mu\sigma}\right].
\end{eqnarray}

\section{Functions of the linearised action}
\label{ap2}

In this Appendix one can find the explicit form of the functions that compose the linearised action.

\begin{eqnarray}
a\left(\Box\right)&=&1-\frac{1}{2}\tilde{F}_{3}\left(\Box\right)\Box-\frac{1}{2}\tilde{F}_{10}\left(\Box\right)\Box^{2}-\frac{1}{2}\tilde{F}_{12}\left(\Box\right)\Box^{2}-2\tilde{F}_{14}\left(\Box\right)\Box
\nonumber
\\
&-&\frac{1}{2}\tilde{F}_{16}\left(\Box\right)\Box^{2}-\frac{1}{2}\tilde{F}_{18}\left(\Box\right)\Box^{2}-\frac{1}{2}\tilde{F}_{20}\left(\Box\right)\Box^{3}-\frac{1}{2}\tilde{F}_{22}\left(\Box\right)\Box^{3}-\tilde{F}_{47}\left(\Box\right)\Box,
\end{eqnarray}

\begin{eqnarray}
b\left(\Box\right)&=&-1+\frac{1}{2}\tilde{F}_{3}\left(\Box\right)\Box+\frac{1}{2}\tilde{F}_{10}\left(\Box\right)\Box^{2}+\frac{1}{2}\tilde{F}_{12}\left(\Box\right)\Box^{2}+2\tilde{F}_{14}\left(\Box\right)\Box
\nonumber
\\
&+&\frac{1}{2}\tilde{F}_{16}\left(\Box\right)\Box^{2}+\frac{1}{2}\tilde{F}_{18}\left(\Box\right)\Box^{2}+\frac{1}{2}\tilde{F}_{20}\left(\Box\right)\Box^{3}+\frac{1}{2}\tilde{F}_{22}\left(\Box\right)\Box^{3}+\tilde{F}_{47}\left(\Box\right)\Box,
\end{eqnarray}

\begin{equation}
c\left(\Box\right)=1+2\tilde{F}_{1}\left(\Box\right)\Box+\tilde{F}_{2}\left(\Box\right)\Box^{2}+\frac{1}{2}\tilde{F}_{3}\left(\Box\right)\Box+\frac{1}{2}\tilde{F}_{5}\left(\Box\right)\Box^{2}+\frac{1}{2}\tilde{F}_{7}\left(\Box\right)\Box^{2}+\frac{1}{2}\tilde{F}_{9}\left(\Box\right)\Box^{3},
\end{equation}

\begin{equation}
d\left(\Box\right)=-1-2\tilde{F}_{1}\left(\Box\right)\Box-\tilde{F}_{2}\left(\Box\right)\Box^{2}-\frac{1}{2}\tilde{F}_{3}\left(\Box\right)\Box-\frac{1}{2}\tilde{F}_{5}\left(\Box\right)\Box^{2}-\frac{1}{2}\tilde{F}_{7}\left(\Box\right)\Box^{2}-\frac{1}{2}\tilde{F}_{9}\left(\Box\right)\Box^{3},
\end{equation}

\begin{eqnarray}
f\left(\Box\right)&=&-\tilde{F}_{1}\left(\Box\right)\Box-\frac{1}{2}\tilde{F}_{2}\left(\Box\right)\Box^{2}-\frac{1}{2}\tilde{F}_{3}\left(\Box\right)\Box-\frac{1}{4}\tilde{F}_{5}\left(\Box\right)\Box^{2}-\frac{1}{4}\tilde{F}_{7}\left(\Box\right)\Box^{2}-\frac{1}{4}\tilde{F}_{9}\left(\Box\right)\Box^{3}-\frac{1}{4}\tilde{F}_{10}\left(\Box\right)\Box^{2}-\frac{1}{4}\tilde{F}_{12}\left(\Box\right)\Box^{2}
\nonumber
\\
&-&\tilde{F}_{14}\left(\Box\right)\Box-\frac{1}{4}\tilde{F}_{16}\left(\Box\right)\Box^{2}-\frac{1}{4}\tilde{F}_{18}\left(\Box\right)\Box^{2}-\frac{1}{4}\tilde{F}_{20}\left(\Box\right)\Box^{3}-\frac{1}{4}\tilde{F}_{22}\left(\Box\right)\Box^{3}-\frac{1}{2}\tilde{F}_{47}\left(\Box\right)\Box,
\end{eqnarray}

\begin{equation}
u\left(\Box\right)=-4\tilde{F}_{1}\left(\Box\right)-\tilde{F}_{5}\left(\Box\right)\Box-\tilde{F}_{7}\left(\Box\right)\Box-\tilde{F}_{9}\left(\Box\right)\Box^{2}+\frac{1}{2}\tilde{F}_{39}\left(\Box\right)+\tilde{F}_{41}\left(\Box\right),
\end{equation}

\begin{equation}
v_{1}\left(\Box\right)=4\tilde{F}_{1}\left(\Box\right)+\tilde{F}_{5}\left(\Box\right)\Box+\tilde{F}_{7}\left(\Box\right)\Box+\tilde{F}_{9}\left(\Box\right)\Box^{2}-\frac{1}{2}\tilde{F}_{39}\left(\Box\right)-\tilde{F}_{41}\left(\Box\right),
\end{equation}

\begin{eqnarray}
v_{2}\left(\Box\right)&=&-\frac{1}{2}\tilde{F}_{3}\left(\Box\right)-\tilde{F}_{10}\left(\Box\right)\Box-\tilde{F}_{12}\left(\Box\right)\Box+\tilde{F}_{9}\left(\Box\right)\Box^{2}-4\tilde{F}_{14}\left(\Box\right)-\tilde{F}_{16}\left(\Box\right)\Box-\tilde{F}_{18}\left(\Box\right)\Box-\tilde{F}_{20}\left(\Box\right)\Box^{2}
\nonumber
\\
&-&\tilde{F}_{22}\left(\Box\right)\Box^{2}+\frac{1}{2}\tilde{F}_{34}\left(\Box\right)+\frac{1}{2}\tilde{F}_{35}\left(\Box\right)+\frac{1}{2}\tilde{F}_{36}\left(\Box\right)+\frac{1}{2}\tilde{F}_{42}\left(\Box\right)-2\tilde{F}_{47}\left(\Box\right),
\end{eqnarray}

\begin{eqnarray}
w\left(\Box\right)&=&-\frac{1}{2}\tilde{F}_{3}\left(\Box\right)-\tilde{F}_{10}\left(\Box\right)\Box-\tilde{F}_{12}\left(\Box\right)\Box+\tilde{F}_{9}\left(\Box\right)\Box^{2}-4\tilde{F}_{14}\left(\Box\right)-\tilde{F}_{16}\left(\Box\right)\Box-\tilde{F}_{18}\left(\Box\right)\Box-\tilde{F}_{20}\left(\Box\right)\Box^{2}
\nonumber
\\
&-&\tilde{F}_{22}\left(\Box\right)\Box^{2}+\frac{1}{2}\tilde{F}_{34}\left(\Box\right)+\frac{1}{2}\tilde{F}_{35}\left(\Box\right)+\frac{1}{2}\tilde{F}_{36}\left(\Box\right)+\frac{1}{2}\tilde{F}_{42}\left(\Box\right)-2\tilde{F}_{47}\left(\Box\right),
\end{eqnarray}

\begin{eqnarray}
q_{1}\left(\Box\right)&=&\frac{1}{2}\tilde{F}_{3}\left(\Box\right)+\frac{1}{2}\tilde{F}_{4}\left(\Box\right)+\frac{1}{2}\tilde{F}_{10}\left(\Box\right)\Box+\frac{1}{2}\tilde{F}_{11}\left(\Box\right)\Box+\frac{1}{2}\tilde{F}_{12}\left(\Box\right)\Box+\frac{1}{2}\tilde{F}_{13}\left(\Box\right)\Box
\nonumber
\\
&+&\frac{1}{2}\tilde{F}_{16}\left(\Box\right)\Box+\frac{1}{2}\tilde{F}_{18}\left(\Box\right)\Box+\frac{1}{2}\tilde{F}_{19}\left(\Box\right)\Box+\frac{1}{2}\tilde{F}_{20}\left(\Box\right)\Box+\frac{1}{2}\tilde{F}_{21}\left(\Box\right)\Box+\frac{1}{2}\tilde{F}_{22}\left(\Box\right)\Box+\frac{1}{2}\tilde{F}_{23}\left(\Box\right)\Box
\nonumber
\\
&+&\tilde{F}_{27}\left(\Box\right)-\frac{1}{2}\tilde{F}_{36}\left(\Box\right)-\frac{1}{2}\tilde{F}_{37}\left(\Box\right)-\frac{1}{2}\tilde{F}_{42}\left(\Box\right)\Box-\frac{1}{2}\tilde{F}_{43}\left(\Box\right)\Box-\tilde{F}_{47}\left(\Box\right),
\end{eqnarray}

\begin{eqnarray}
q_{2}\left(\Box\right)&=&\frac{1}{2}\tilde{F}_{3}\left(\Box\right)-\frac{1}{2}\tilde{F}_{4}\left(\Box\right)+\frac{1}{2}\tilde{F}_{10}\left(\Box\right)\Box-\frac{1}{2}\tilde{F}_{11}\left(\Box\right)\Box+\frac{1}{2}\tilde{F}_{12}\left(\Box\right)\Box-\frac{1}{2}\tilde{F}_{13}\left(\Box\right)\Box+2\tilde{F}_{14}\left(\Box\right)-2\tilde{F}_{15}\left(\Box\right)
\nonumber
\\
&+&\frac{1}{2}\tilde{F}_{16}\left(\Box\right)\Box+\frac{1}{2}\tilde{F}_{20}\left(\Box\right)\Box-\frac{1}{2}\tilde{F}_{21}\left(\Box\right)\Box+\frac{1}{2}\tilde{F}_{22}\left(\Box\right)\Box-\frac{1}{2}\tilde{F}_{23}\left(\Box\right)\Box+\tilde{F}_{28}\left(\Box\right)-\frac{1}{2}\tilde{F}_{36}\left(\Box\right)+\frac{1}{2}\tilde{F}_{37}\left(\Box\right)
\nonumber
\\
&-&\frac{1}{2}\tilde{F}_{42}\left(\Box\right)\Box+\frac{1}{2}\tilde{F}_{43}\left(\Box\right)\Box,
\end{eqnarray}

\begin{equation}
q_{3}\left(\Box\right)=-\tilde{F}_{17}\left(\Box\right)\Box-\tilde{F}_{18}\left(\Box\right)\Box+\tilde{F}_{19}\left(\Box\right)\Box+\tilde{F}_{29}\left(\Box\right)+\tilde{F}_{34}\left(\Box\right)-\tilde{F}_{35}\left(\Box\right)-\tilde{F}_{38}\left(\Box\right)-\tilde{F}_{44}\left(\Box\right)\Box+2\tilde{F}_{47}\left(\Box\right),
\end{equation}

\begin{equation}
q_{4}\left(\Box\right)=-\tilde{F}_{14}\left(\Box\right)-\tilde{F}_{15}\left(\Box\right)+\tilde{F}_{30}\left(\Box\right),
\end{equation}

\begin{equation}
q_{5}\left(\Box\right)=4\tilde{F}_{1}\left(\Box\right)+2\tilde{F}_{2}\left(\Box\right)\Box+\frac{1}{2}\tilde{F}_{3}\left(\Box\right)-\frac{1}{2}\tilde{F}_{4}\left(\Box\right)+\tilde{F}_{5}\left(\Box\right)\Box+\tilde{F}_{7}\left(\Box\right)\Box+\tilde{F}_{9}\left(\Box\right)\Box^{2}+\tilde{F}_{31}\left(\Box\right)-\frac{1}{2}\tilde{F}_{39}\left(\Box\right)-\frac{1}{2}\tilde{F}_{40}\left(\Box\right)-2\tilde{F}_{41}\left(\Box\right),
\end{equation}

\begin{equation}
q_{6}\left(\Box\right)=\tilde{F}_{3}\left(\Box\right)+\tilde{F}_{4}\left(\Box\right)+\tilde{F}_{32}\left(\Box\right)+\frac{1}{2}\tilde{F}_{36}\left(\Box\right)+\frac{1}{2}\tilde{F}_{37}\left(\Box\right)-\tilde{F}_{38}\left(\Box\right)-\frac{1}{2}\tilde{F}_{39}\left(\Box\right)+\frac{1}{2}\tilde{F}_{40}\left(\Box\right)+\frac{1}{2}\tilde{F}_{45}\left(\Box\right)\Box+\frac{1}{2}\tilde{F}_{46}\left(\Box\right)\Box,
\end{equation}

\begin{equation}
p_{1}\left(\Box\right)=\tilde{F}_{14}\left(\Box\right)\Box+\tilde{F}_{15}\left(\Box\right)\Box+\tilde{F}_{24}\left(\Box\right),
\end{equation}

\begin{equation}
p_{2}\left(\Box\right)=\frac{1}{2}\tilde{F}_{18}\left(\Box\right)\Box^{2}-\frac{1}{2}\tilde{F}_{19}\left(\Box\right)\Box^{2}+\tilde{F}_{25}\left(\Box\right)+\tilde{F}_{34}\left(\Box\right)-\tilde{F}_{47}\left(\Box\right)\Box,
\end{equation}

\begin{equation}
p_{3}\left(\Box\right)=\frac{1}{2}\tilde{F}_{3}\left(\Box\right)\Box+\frac{1}{2}\tilde{F}_{4}\left(\Box\right)\Box+\tilde{F}_{26}\left(\Box\right)-\frac{1}{2}\tilde{F}_{39}\left(\Box\right)\Box+\frac{1}{2}\tilde{F}_{40}\left(\Box\right)\Box,
\end{equation}

\begin{eqnarray}
s\left(\Box\right)&=&-\frac{1}{2}\tilde{F}_{10}\left(\Box\right)-\frac{1}{2}\tilde{F}_{11}\left(\Box\right)-\frac{1}{2}\tilde{F}_{12}\left(\Box\right)-\frac{1}{2}\tilde{F}_{13}\left(\Box\right)-\frac{1}{2}\tilde{F}_{16}\left(\Box\right)+\tilde{F}_{17}\left(\Box\right)-\frac{1}{2}\tilde{F}_{20}\left(\Box\right)-\frac{1}{2}\tilde{F}_{21}\left(\Box\right)-\frac{1}{2}\tilde{F}_{22}\left(\Box\right)
\nonumber
\\
&-&\frac{1}{2}\tilde{F}_{23}\left(\Box\right)+\tilde{F}_{33}\left(\Box\right)+\frac{1}{2}\tilde{F}_{42}\left(\Box\right)-\frac{1}{2}\tilde{F}_{43}\left(\Box\right)-\frac{1}{2}\tilde{F}_{44}\left(\Box\right)-\frac{1}{2}\tilde{F}_{45}\left(\Box\right)-\frac{1}{2}\tilde{F}_{46}\left(\Box\right).
\end{eqnarray}

\section{Poincar\'e Gauge Gravity as the local limit}
\label{ap3}

Here we will give more insight on how Poincar\'e Gauge Gravity can be recasted as the local limit of our theory.

Poincar\'e Gauge Gravity is constructed by gauging the Poincar\'e group, that is formed of the homogeneous Lorentz group $SO(3,1)$ plus the spacetime translations. The field strenght of the latter is the torsion field, while the Riemmannian curvature as the field strength of translations, while the Riemann curvature is associated to the homogeneous part \cite{Gauge}.
Inspired by Yang-Mills theories, the usual Lagrangian of this theory is built using quadratic terms in the field strengths, namely\footnote{Please note that we have used the contorsion tensor instead of the torsion one without losing any generality, since they are related by a linear expression.}
\begin{eqnarray}
\label{9lag}
\mathcal{L}_{\rm PG}&=&\tilde{R}+b_{1}\tilde{R}^{2}+b_{2}\tilde{R}_{\mu\nu\rho\sigma}\tilde{R}^{\mu\nu\rho\sigma}+b_{3}\tilde{R}_{\mu\nu\rho\sigma}\tilde{R}^{\rho\sigma\mu\nu}+b_{4}\tilde{R}_{\mu\nu\rho\sigma}\tilde{R}^{\mu\rho\nu\sigma}+b_{5}\tilde{R}_{\mu\nu}\tilde{R}^{\mu\nu}+b_{6}\tilde{R}_{\mu\nu}\tilde{R}^{\nu\mu}
\nonumber
\\
&&+a_{1}K_{\mu\nu\rho}K^{\mu\nu\rho}+a_{2}K_{\mu\nu\rho}K^{\mu\rho\nu}+a_{3}K_{\nu\,\,\,\,\,\mu}^{\,\,\,\mu}K_{\,\,\,\,\,\,\rho}^{\nu\rho},
\end{eqnarray}
which is usually known as the nine-parameter Lagrangian. Since in the torsion-free limit we want to recover the results of usual IDG, the local limit at zero torsion must be GR. This fact imposes the following constraints in the Lagrangian \eqref{9lag}
\begin{eqnarray}
\label{Bonnet}
b_{6}=-4b_{1}-b_{5}\;,\;b_{4}=2\left(b_{1}-b_{2}-b_{3}\right),
\end{eqnarray}
where we have used the topological character of the Gauss-Bonnet term.\\
From \eqref{9lag}, and taking into account \eqref{Bonnet}, one can calculate the linearised Lagrangian just by substituing the expressions of the curvature tensors (\ref{riemann},\ref{ricci},\ref{scalar}), obtaining
\begin{eqnarray}
\label{PGlinear}
\mathcal{L}_{{\rm PG}}^{{\rm linear}}&=&\frac{1}{2}h_{\mu\nu}\Box h^{\mu\nu}-h_{\mu}^{\,\,\alpha}\partial_{\alpha}\partial_{\sigma}h^{\sigma\mu}+h\partial_{\mu}\partial_{\nu}h^{\mu\nu}-\frac{1}{2}h\Box h-4b_{1}h\Box\partial_{\rho}K_{\,\,\,\,\,\sigma}^{\rho\sigma}+4b_{1}h_{\mu\nu}\partial^{\mu}\partial^{\nu}\partial_{\rho}K_{\,\,\,\,\,\sigma}^{\rho\sigma}
\nonumber
\\
&&-\left(6b_{1}+b_{5}\right)h_{\mu\nu}\partial^{\nu}\partial_{\sigma}\partial_{\rho}K^{\mu\sigma\rho}-\left(6b_{1}+b_{5}\right)h_{\mu\nu}\Box\partial_{\rho}K^{\rho\mu\nu}+K^{\mu\sigma\lambda}\left(a_{1}+2b_{2}\Box\right)K_{\mu\sigma\lambda}
\nonumber
\\
&&+K^{\mu\sigma\lambda}\left(a_{2}-2\left(b_{1}-b_{2}-b_{3}\right)\Box\right)K_{\mu\lambda\sigma}+K_{\mu\,\,\rho}^{\,\,\rho}\left(a_{3}+b_{5}\Box\right)K_{\,\,\,\,\,\sigma}^{\mu\sigma}+\left(b_{5}-2b_{1}+2b_{2}+2b_{3}\right)K_{\,\,\nu\rho}^{\mu}\partial_{\mu}\partial_{\sigma}K^{\sigma\nu\rho}
\nonumber
\\
&&+\left(-4b_{1}-b_{5}+4b_{3}\right)K_{\,\,\nu\rho}^{\mu}\partial_{\mu}\partial_{\sigma}K^{\sigma\rho\nu}+4\left(b_{1}-b_{2}-b_{3}\right)K_{\mu\,\,\,\,\,\nu}^{\,\,\rho}\partial_{\rho}\partial_{\sigma}K^{\mu\nu\sigma}-2b_{2}K_{\mu\,\,\,\,\,\nu}^{\,\,\rho}\partial_{\rho}\partial_{\sigma}K^{\mu\sigma\nu}
\nonumber
\\
&&+\left(4b_{1}+b_{3}\right)K_{\,\,\,\,\,\rho}^{\mu\rho}\partial_{\mu}\partial_{\nu}K_{\,\,\,\,\,\sigma}^{\nu\sigma}+2b_{5}K_{\,\,\,\lambda\sigma}^{\lambda}\partial_{\mu}\partial_{\alpha}K^{\sigma\mu\alpha}.
\end{eqnarray}
At the same time, the local limit of our theory can be expressed, applying the constraints \eqref{constraints}, as
\begin{eqnarray}
\label{linearlocal}
\mathcal{L}\left(M_{S}\rightarrow\infty\right)&=&\frac{1}{2}a\left(0\right)h_{\mu\nu}\Box h^{\mu\nu}-a\left(0\right)h_{\mu}^{\,\,\alpha}\partial_{\alpha}\partial_{\sigma}h^{\sigma\mu}+c\left(0\right)h\partial_{\mu}\partial_{\nu}h^{\mu\nu}-\frac{1}{2}c\left(0\right)h\Box h+\frac{a\left(0\right)-c\left(0\right)}{\Box}h^{\lambda\sigma}\partial_{\sigma}\partial_{\lambda}\partial_{\mu}\partial_{\nu}h^{\mu\nu}
\nonumber
\\
&&+u\left(0\right)h\Box\partial_{\rho}K_{\,\,\,\,\,\sigma}^{\rho\sigma}-u\left(0\right)h_{\mu\nu}\partial^{\mu}\partial^{\nu}\partial_{\rho}K_{\,\,\,\,\,\sigma}^{\rho\sigma}+v_{2}\left(0\right)h_{\mu\nu}\partial^{\nu}\partial_{\sigma}\partial_{\rho}K^{\mu\sigma\rho}+v_{2}\left(0\right)h_{\mu\nu}\Box\partial_{\rho}K^{\rho\mu\nu}
\nonumber
\\
&&+p_{1}\left(0\right)K^{\mu\sigma\lambda}K_{\mu\sigma\lambda}+p_{2}\left(0\right)K^{\mu\sigma\lambda}K_{\mu\lambda\sigma}+p_{3}\left(0\right)K_{\mu\,\,\rho}^{\,\,\rho}K_{\,\,\,\,\,\sigma}^{\mu\sigma}+q_{1}\left(0\right)K_{\,\,\nu\rho}^{\mu}\partial_{\mu}\partial_{\sigma}K^{\sigma\nu\rho}
\nonumber
\\
&&+q_{2}\left(0\right)K_{\,\,\nu\rho}^{\mu}\partial_{\mu}\partial_{\sigma}K^{\sigma\rho\nu}+q_{3}\left(0\right)K_{\mu\,\,\,\,\,\nu}^{\,\,\rho}\partial_{\rho}\partial_{\sigma}K^{\mu\nu\sigma}+q_{4}\left(0\right)K_{\mu\,\,\,\,\,\nu}^{\,\,\rho}\partial_{\rho}\partial_{\sigma}K^{\mu\sigma\nu}+q_{5}\left(0\right)K_{\,\,\,\,\,\rho}^{\mu\rho}\partial_{\mu}\partial_{\nu}K_{\,\,\,\,\,\sigma}^{\nu\sigma}
\nonumber
\\
&&+q_{6}\left(0\right)K_{\,\,\,\lambda\sigma}^{\lambda}\partial_{\mu}\partial_{\alpha}K^{\sigma\mu\alpha}+s\left(0\right)K_{\mu}^{\,\,\nu\rho}\partial_{\nu}\partial_{\rho}\partial_{\alpha}\partial_{\sigma}K^{\mu\alpha\sigma}.
\end{eqnarray} 
It is straightforward to realise that we have more free parameters in \eqref{linearlocal} than in \eqref{PGlinear}, which means that if we want \eqref{PGlinear} as the local limit, we will need to impose more constraints in the parameters in \eqref{linearlocal}. The question now is if there exists a PG theory that can be recasted as the local limit of our theory without compromising the independence of the parameters. The answer is affirmative, as can be seen in the following Lagrangian
\begin{eqnarray}
\label{genPG}
\mathcal{L}_{{\rm GPG}}&=&\tilde{R}+b_{1}\tilde{R}^{2}+b_{2}\tilde{R}_{\mu\nu\rho\sigma}\tilde{R}^{\mu\nu\rho\sigma}+b_{3}\tilde{R}_{\mu\nu\rho\sigma}\tilde{R}^{\rho\sigma\mu\nu}+2\left(b_{1}-b_{2}-b_{3}\right)\tilde{R}_{\mu\nu\rho\sigma}\tilde{R}^{\mu\rho\nu\sigma}+b_{5}\tilde{R}_{\mu\nu}\tilde{R}^{\mu\nu}
\nonumber
\\
&&-\left(4b_{1}+b_{5}\right)\tilde{R}_{\mu\nu}\tilde{R}^{\nu\mu}+a_{1}K_{\mu\nu\rho}K^{\mu\nu\rho}+a_{2}K_{\mu\nu\rho}K^{\mu\rho\nu}+a_{3}K_{\nu\,\,\,\,\,\mu}^{\,\,\,\mu}K_{\,\,\,\,\,\,\rho}^{\nu\rho}+c_{1}K_{\,\,\nu\rho}^{\mu}\nabla_{\mu}\nabla_{\sigma}K^{\sigma\nu\rho}
\nonumber
\\
&&+c_{2}K_{\,\,\nu\rho}^{\mu}\nabla_{\mu}\nabla_{\sigma}K^{\sigma\rho\nu}+c_{3}K_{\mu\,\,\,\,\,\nu}^{\,\,\rho}\nabla_{\rho}\nabla_{\sigma}K^{\mu\nu\sigma}+c_{4}K_{\mu\,\,\,\,\,\nu}^{\,\,\rho}\nabla_{\rho}\nabla_{\sigma}K^{\mu\sigma\nu},
\end{eqnarray}
which is Poincar\'e Gauge invariant and local. Its corresponding linearised expression is
\begin{eqnarray}
\label{genPGlin}
\mathcal{L}_{{\rm GPG}}^{{\rm linear}}&=&\frac{1}{2}h_{\mu\nu}\Box h^{\mu\nu}-h_{\mu}^{\,\,\alpha}\partial_{\alpha}\partial_{\sigma}h^{\sigma\mu}+h\partial_{\mu}\partial_{\nu}h^{\mu\nu}-\frac{1}{2}h\Box h-4b_{1}h\Box\partial_{\rho}K_{\,\,\,\,\,\sigma}^{\rho\sigma}+4b_{1}h_{\mu\nu}\partial^{\mu}\partial^{\nu}\partial_{\rho}K_{\,\,\,\,\,\sigma}^{\rho\sigma}
\nonumber
\\
&&-\left(6b_{1}+b_{5}\right)h_{\mu\nu}\partial^{\nu}\partial_{\sigma}\partial_{\rho}K^{\mu\sigma\rho}-\left(6b_{1}+b_{5}\right)h_{\mu\nu}\Box\partial_{\rho}K^{\rho\mu\nu}+K^{\mu\sigma\lambda}\left(a_{1}+2b_{2}\Box\right)K_{\mu\sigma\lambda}
\nonumber
\\
&&+K^{\mu\sigma\lambda}\left(a_{2}-2\left(b_{1}-b_{2}-b_{3}\right)\Box\right)K_{\mu\lambda\sigma}+K_{\mu\,\,\rho}^{\,\,\rho}\left(a_{3}+b_{5}\Box\right)K_{\,\,\,\,\,\sigma}^{\mu\sigma}+\left(b_{5}-2b_{1}+2b_{2}+2b_{3}+c_{1}\right)K_{\,\,\nu\rho}^{\mu}\partial_{\mu}\partial_{\sigma}K^{\sigma\nu\rho}
\nonumber
\\
&&+\left(-4b_{1}-b_{5}+4b_{3}+c_{2}\right)K_{\,\,\nu\rho}^{\mu}\partial_{\mu}\partial_{\sigma}K^{\sigma\rho\nu}+\left(4b_{1}-4b_{2}-4b_{3}+c_{3}\right)K_{\mu\,\,\,\,\,\nu}^{\,\,\rho}\partial_{\rho}\partial_{\sigma}K^{\mu\nu\sigma}
\nonumber
\\
&&-\left(2b_{2}-c_{4}\right)K_{\mu\,\,\,\,\,\nu}^{\,\,\rho}\partial_{\rho}\partial_{\sigma}K^{\mu\sigma\nu}+\left(4b_{1}+b_{3}\right)K_{\,\,\,\,\,\rho}^{\mu\rho}\partial_{\mu}\partial_{\nu}K_{\,\,\,\,\,\sigma}^{\nu\sigma}+2b_{5}K_{\,\,\,\lambda\sigma}^{\lambda}\partial_{\mu}\partial_{\alpha}K^{\sigma\mu\alpha}.
\end {eqnarray}
Therefore, one finds the following relations for the local limit of the functions involved in the linearised action \eqref{linearlocal} and the parameters in \eqref{genPGlin}
\begin{eqnarray}
\label{localconditions}
&&a\left(0\right)=1,\;c\left(0\right)=1,\;u\left(0\right)=-4b_{1},\;v_{2}\left(0\right)=-4\left(6b_{1}+b_{5}\right),\;p_{1}\left(0\right)=a_{1}+2b_{2}\Box,\;p_{2}\left(0\right)=a_{2}-2\left(b_{1}-b_{2}-b_{3}\right)\Box,\;
\nonumber
\\
&&p_{3}\left(0\right)=a_{3}+b_{5}\Box,\;q_{1}\left(0\right)=b_{5}-2b_{1}+2b_{2}+2b_{3}+c_{1},\;q_{2}\left(0\right)=-4b_{1}-b_{5}+4b_{3}+c_{2},\;q_{3}\left(0\right)=4b_{1}-4b_{2}-4b_{3}+c_{3},\;
\nonumber
\\
&&q_{4}\left(0\right)=-2b_{2}+c_{4},\;q_{5}\left(0\right)=4b_{1}+b_{3},\;q_{6}\left(0\right)=2b_{5},\;s\left(0\right)=0.
\end{eqnarray}
It can be observe that these limits do not impose new relations between the functions. \\
Hence, we have proved that if the previous limits apply, the local limit of our theory is a local PG theory, concretly the one described by the Lagrangian \eqref{genPG}.

\bibliographystyle{JHEP}
\bibliography{References-1}

\end{document}